\documentclass[11pt,a4paper]{article}            
               
\usepackage[skins,theorems]{tcolorbox}
\tcbset{highlight math style={enhanced,
  colframe=red,colback=white,arc=0pt,boxrule=1pt}}
\usepackage[bookmarksopen, bookmarksnumbered, bookmarksopenlevel=2]{hyperref}
\usepackage{tikz}
\usepackage{tikz-3dplot}
\usetikzlibrary{calc}
\usetikzlibrary{decorations} %
\usepackage[UKenglish]{babel}
\usepackage[toc,page]{appendix}
\usepackage{amsmath}
\usepackage{amssymb}
\usepackage{graphicx}
\usepackage{hhline}
\usepackage[bf]{caption}
\usepackage{cite}
\usepackage[vcentermath]{youngtab}
\usepackage{geometry}
\usepackage{slashed}
\usepackage{color}
\usepackage{stackrel}
\usepackage{tikz}
\usepackage{tikz-cd} 
\usepackage{empheq}

\newcommand{\bqa}{\begin{eqnarray}}
\newcommand{\eqa}{\end{eqnarray}}

\newcommand{\nn}{\nonumber}

\hypersetup{
    pdftitle={},
    pdfauthor={},
    pdfsubject={}
}
\numberwithin{equation}{section}
\numberwithin{table}{section}

\makeatletter

\newcommand{\be}{\begin{equation}}
\newcommand{\ee}{\end{equation}}
\newcommand{\beq}{\begin{equation}}
\newcommand{\eeq}{\end{equation}}
\newcommand{\ba}{\begin{aligned}}
\newcommand{\ea}{\end{aligned}}

\newcommand{\bea}{\begin{eqnarray}}
\newcommand{\eea}{\end{eqnarray}}

\newcommand{\cO}{\mathcal{O}}

\newcommand{\Ub}{{b}}
\newcommand{\ik}{{\kappa}}

\def\tr{\mathop{\mathrm{tr}}\nolimits}

\def\unit{{1\kern-.65ex {\rm l}}}
\def\1{{1\kern-.65ex {\rm l}}}

\def\IZ{\mathbb{Z}}
\def\IP{\mathbb{P}}

\newcount\hour \newcount\minute
\hour=\time \divide \hour by 60
\minute=\time
\def\now{%
\ifnum \hour<13
  \ifnum \hour=0 \advance \hour by 12 \number\hour:\else \number\hour:\fi%
     \ifnum \minute<10 0\fi%
     \number\minute%
\ A.M.%
\else \advance \hour by -12 \number\hour:%
  \ifnum \minute<10 0\fi%
  \number\minute%
  \ P.M.%
\fi%
}

\def\fnote#1#2{\begingroup\def\thefootnote{#1}\footnote{#2}
     \addtocounter{footnote}{-1}\endgroup}

\makeatother

\begin{document}

\begin{flushright}
{\tt\normalsize CERN-TH-2018-054}\\
\end{flushright}

\vskip 40 pt
\begin{center}
{\huge \bf 6d SCFTs and U(1) Flavour Symmetries}

\vskip 7 mm

Seung-Joo Lee${}^{1}$, 
Diego Regalado${}^{1}$ 
and Timo Weigand${}^{1,2}$

\vskip 7 mm

\small ${}^{1}${\it CERN, Theory Division, \\ CH-1211 Geneva 23, Switzerland} \\[3 mm]
\small ${}^{2}${\it Institut f\"ur Theoretische Physik, Ruprecht-Karls-Universit\"at, \\
Philosophenweg 19, 69120 Heidelberg, Germany}

\fnote{}{seung.joo.lee@cern.ch, diego.regalado@cern.ch, timo.weigand@cern.ch}

\end{center}

\vskip 7mm

\begin{abstract}

We study the behaviour of abelian gauge symmetries in six-dimensional $N=(1,0)$ theories upon decoupling gravity and investigate abelian flavour symmetries in
the context of 6d $N=(1,0)$ SCFTs.
From a supergravity perspective, the anomaly cancellation mechanism implies that abelian gauge symmetries can only survive as global symmetries as gravity is decoupled.
The flavour symmetries obtained in this way are shown to be free of ABJ anomalies, and their 't Hooft anomaly polynomial in the decoupling limit is obtained explicitly.
In an F-theory realisation the decoupling of abelian gauge symmetries implies that a mathematical object known as the height pairing of a rational section is
not contractible as a curve on the base of an elliptic Calabi-Yau threefold. We prove this prediction from supergravity by making use of the properties of the Mordell-Weil group
of rational sections. In the second part of this paper we study the appearance of abelian flavour symmetries in 6d $N=(1,0)$ SCFTs.
We elucidate both the geometric origin of such flavour symmetries in F-theory and their field theoretic interpretation in terms of
suitable linear combinations of geometrically massive $U(1)$s. Our general results are illustrated in various explicit examples.

\end{abstract}

\vfill

\thispagestyle{empty}
\setcounter{page}{0}
\newpage

\tableofcontents



\section{Introduction}

Abelian gauge theories enjoy a somewhat special status as quantum field theories compared to their non-abelian cousins. 
In four dimensions, their perturbative beta-function is non-negative and the theory, once coupled to matter, is bound to leave the perturbative regime at very high energies. Extrapolating the behaviour of the gauge coupling towards the ultra-violet (UV), it is natural to speculate that a proper definition of the theory in the UV requires new degrees of freedom. One possibility is an embedding into an asymptotically free, and hence ultra-violet complete, non-abelian gauge theory above certain energy scales.
In absence of such a protection mechanism, quantum gravity effects might be required to render the theory well-defined. 

In six dimensions, on the other hand, the fate of being non UV-complete by themselves is shared not only by abelian, but even by non-abelian gauge theories. Indeed, all gauge theories are non-renormalisable and hence become strongly coupled towards the UV. Interestingly, in presence of eight supercharges, the degrees of freedom needed to render a non-abelian gauge theory ultra-violet complete are not necessarily gravitational in nature. Rather, an $N=(1,0)$ supersymmetric gauge theory contains anti-self-dual tensor fields coupling to string-like objects in six dimensions. The potentially dangerous strong coupling limit of the gauge theory coincides with the limit in which these strings become light and eventually tensionless, furnishing infinitely many new degrees of freedom which enter the dynamics of the theory \cite{Seiberg:1996vs,Witten:1995zh}. In the case of a non-abelian gauge theory this leads to a conjectured non-trivial UV fixed point dubbed $N=(1,0)$ superconformal field theory (SCFT), which can exist even without coupling in truly gravitational degrees of freedom.

Given this powerful UV protection mechanism, it is natural to wonder about the fate of six-dimensional abelian gauge theories as we approach strong coupling. Despite little explicit knowledge about the microscopic physics of the 6d $N=(1,0)$ SCFTs, it is widely believed that the existence of a UV fixed point is intimately related to the non-abelian nature of the gauge and tensor theory to which the tensionless strings are coupled.
If this is the case, an abelian theory, unless embedded into a non-abelian model at high energies, must be UV completed by gravity itself, much as in four dimensions. 

In this article we show that this natural assumption is indeed correct both from the perspective of a 6d $N=(1,0)$ supergravity analysis and in the context of string theory, and analyse its consequences for the possibility of having abelian decorations of $N=(1,0)$ SCFTs. 
As we will see in \autoref{sec_U1gaugsymmetries}, the well-known structure of anomaly cancellation alone implies that an abelian gauge theory in six dimensions cannot exist in absence of gravity. More precisely, as we decouple the gravitational multiplet by taking the 6d Planck mass to infinity, the celebrated Green-Schwarz-Sagnotti-West  mechanism \cite{Green:1984bx,Sagnotti:1992qw} (GS for short) fails to cancel the 1-loop abelian gauge anomalies after we discard the contribution from the gravi-tensor. As a result the abelian theory on its own is inconsistent as a gauge theory. The situation is strikingly different from its non-abelian counterpart, where the GS mechanism for the cancellation of the pure gauge anomalies is not necessarily affected by the decoupling of the gravi-tensor. In this sense, abelian gauge theories coupled to matter do not exist in six dimensions unless coupled to gravity (or embedded in a non-abelian theory). In particular, there is no UV protection mechanism in the form of an $N=(1,0)$ SCFT, as expected on general grounds.  Rather, as we decouple gravity, the abelian gauge coupling necessarily tends to zero as well. In this field theory limit, the abelian symmetry remains only as a global, or flavour, symmetry. While the decoupling of the gravi-tensor is fatal to the cancellation of purely abelian gauge anomalies, the mixed abelian-non-abelian anomalies are not affected (provided the non-abelian gauge theory fields themselves stay dynamical  in the field theory limit). 
The $U(1)$ flavour symmetry is hence free of chiral, or ABJ, anomalies, while the remnant of the uncancelled $U(1)$ gauge and mixed gauge-gravitational anomalies gives rise to non-trivial 't Hooft anomalies.

This simple observation manifests itself in a beautiful manner in the context of a string-theoretic realisation of such theories. 
Indeed, it is well appreciated in the literature that abelian gauge theories have a rather special status also concerning their explicit origin in string compactifications. 
In the context of brane constructions, non-abelian gauge symmetries are locally supported on individual brane stacks wrapping suitable cycles of the compactification space.
If we stay, for concreteness, in the framework of Type IIB/F-theory compactifications with 7-branes, the latter wrap complex curves on a 2-complex-dimensional compactification space $B_2$.
Decoupling gravity from non-abelian gauge theories corresponds to taking the volume of $B_2$ to infinity while keeping the volume of the wrapped curve finite. Curves for which this is possible are called contractible, because mathematically this process is equivalent to shrinking the curve to zero size while keeping the base volume finite.

$U(1)$ gauge symmetries, on the other hand, are sensitive to the global details of the compactification space. From the Type II brane perspective, the massless $U(1)$s are linear combinations of $U(1)$ factors from branes at different locations of the compactification space.
In F-theory, this special status of the abelian gauge symmetries is reflected in the fact that they originate in the global rational sections of the underlying elliptic fibration \cite{Morrison:1996pp}. We review this description in some detail in \autoref{reviewMW} in the hope of making this article accessible not only to the F-theory afficionado. The explicit realisation of abelian gauge symmetries in global F-theory compactifications has been studied in detail in the recent F-theory literature, beginning with \cite{Grimm:2010ez}.
Similarly to non-abelian gauge theories, one can attribute to each abelian gauge symmetry a curve class on the base, known as the height pairing associated with the section.
The volume of this curve is proportional to the square of the inverse gauge coupling (at least in the absence of kinetic mixing), and the curve class itself is the anomaly coefficient governing the GS mechanism for the $U(1)$ \cite{Park:2011ji,Morrison:2012ei}.

As a simple observation we point out in \autoref{sec_U1gaugsymmetries} that the mechanism of anomaly cancellation (at least in absence of kinetic mixing) already implies that this curve has positive self-intersection and hence cannot be contractible. As a result, when we take the volume of the compactification space to be infinite, the curve in question acquires infinite volume as well and the gauge theory coupling vanishes.
Given that the height pairing is an object of intensive study in mathematics in its own right, this is an interesting prediction of F-theory for the geometry of rational sections.

In \autoref{sec_DecoupFCY3} we prove this geometric property directly and without the use of anomaly cancellation.
The proof first makes use of a theorem of Kodaira \cite{Kodaira} and N\'eron \cite{Neron} to argue that the height pairing can always be written as the {\it sum} of two {effective} divisors, one of which is the  anti-canonical divisor; this is a generalization of the Cox-Zucker approach \cite{CoxZucker} to studying rational sections on elliptic surfaces. We then proceed to show that the anti-canonical curve is not contractible. Indeed, on bases supporting minimal elliptic fibrations, contractible curves necessarily give rise to singularities of the form $\mathbb C^2/\Gamma$ with $\Gamma$ a discrete subgroup of $U(2)$ \cite{Heckman:2013pva}. As we will discuss, since these singularities are rational, they cannot be blown up into an anti-canonical curve.

Having established, both in supergravity and string/F-theory, that abelian symmetries can only survive as global symmetries after decoupling gravity, we can wonder about their role in the context of 6d $N=(1,0)$ SCFTs, as classified recently in F-theory  \cite{Heckman:2013pva,DelZotto:2014hpa,Heckman:2015bfa,Heckman:2015ola} and from the perspective of field theory \cite{Bhardwaj:2015xxa}. Since the abelian symmetries survive the decoupling limit without acquiring dangerous ABJ anomalies with the dynamical gauge groups, they remain as global symmetries of the non-abelian gauge theories on the tensor branch. 
More precisely, we will
focus in this work on the global symmetries which can be identified via their action on 6d matter hypermultiplets. 
We argue in \autoref{sec_examples} that the global abelian symmetries of this type encountered  in a given F-theory realisation of the decoupling limit can be understood as linear combinations of the `diagonal $U(1)$s' associated with the maximal non-abelian flavor group of the model, possibly with an admixture of Cartan $U(1)$s within this maximal non-abelian flavour group. Field theoretically, the linear combination is determined by requiring the absence of mixed cubic 1-loop anomalies, which would require the $U(1)$s to be `geometrically massive' (as studied in the F-theory context in \cite{Grimm:2011tb}).
In this sense the a priori rich set of different options to realize $U(1)$ gauge symmetries in a globally defined F-theory model reduces to a rather tractable set of possible flavour symmetries after decoupling gravity, which can essentially be determined locally. This is in agreement with the way how detailed global information about the existence of rational sections gets washed out in the local decoupling limit.

The possibility of having abelian flavour symmetries has first appeared in this context in \cite{Heckman:2016ssk} in a field theoretic analysis, starting from a non-abelian flavour symmetry and breaking it by T-brane data to remnants which sometimes contain abelian factors. However, the charges of the representations and the geometric origin of the abelian symmetries  have not been determined. 
Recently, in \cite{Anderson:2018heq} examples of F-theory compactifications have been constructed with a decoration of an $N=(2,0)$ SCFT sector which carries charges, in the supergravity regime, under abelian and discrete gauge symmetries. Given our analysis of \autoref{sec_U1gaugsymmetries}, these theories have abelian (or discrete) flavor symmetries in the decoupling limit of gravity, and, as we expect, also in the SCFT phase. 
In \autoref{sec_examples} we give examples of such theories, proceeding along two complementary approaches.
First, in \autoref{sec_Globalmodel1} and \autoref{sec_Globalmodel2}, we construct globally defined F-theory models over compact base spaces which contain shrinkable curves. Enhancing the non-abelian gauge group over these curves leads to a theory which can be enhanced by abelian gauge symmetries if we constrain the elliptic fibration further such as to admit an extra independent section. We then analyse the limit of decoupling gravity and show explicitly that the resulting abelian flavour symmetry is free of ABJ anomalies. We compute the 't Hooft anomaly polynomial and conjecture that, since this does not change along the tensor branch, the resulting SCFT has an abelian flavour symmetry.

As we explain in \autoref{subsec_fieldexp}, the abelian flavour symmetry obtained in these examples can indeed be understood in terms of a suitable linear combination $U(1)_m$ of individually massive $U(1)$ factors of the maximal field theoretic flavour symmetry. The key point in this interpretation is to assign to $N$ hypermultiplets in a complex representation a maximal flavour group of $U(N) = SU(N) \times U(1)/\mathbb Z_N$. This leads to a natural proposal for the general form in particular of the non-abelian part of the flavour group.
When we take the decoupling limit of a globally consistent model with a $U(1)_A$ gauge group, the action of the latter on the charged states is to be identified with a linear combination of this $U(1)_m$ and some Cartan $U(1)_c$ within the non-abelian part of the flavour group. 
Even though we started, heuristically, from a global model with an abelian gauge group, we stress that once the dust has settled, the abelian flavour symmetry of any local model can be determined fully locally.
Whether or not it survives as a gauge theory in the presence of gravity depends on the chosen global completion.

To exemplify this, in \autoref{sec_localmodel} we turn tables round and ask in which cases a local model {\it might} have an embedding into a global geometry with an abelian gauge group.
We investigate this for the $N=(1,0)$ conformal matter theory based on a chain of $(-1)  - (-3)  - (-1)$ curves enhanced to an $\mathfrak{so}(8)$ gauge symmetry \cite{Heckman:2015bfa}.
The flavour group of the local model only involves $Sp(1)$ factors and hence no extra abelian flavour symmetries. Nonetheless, some linear combinations of the flavour Cartan $U(1)$s might be realized as gauge symmetries in a global completion.
Unfortunately, to date the most general form of an elliptic fibration with an extra section is not known due to subtleties associated with non-unique factorization domains  \cite{Klevers:2014bqa,Klevers:2016jsz,Morrison:2016xkb,Raghuram:2017qut}. To be concrete we require that the model arises as the local limit of an elliptic fibration which can be written in the Morrison-Park form \cite{Morrison:2012ei}. The canonical (and some non-canonical) forms of non-abelian gauge enhancements have been classified in this framework in \cite{Kuntzler:2014ila}. The details of these enhancements, together with anomaly cancellation, constrain the form of the height pairing and hence the possible abelian charges in a potential global completion of this type.

We summarize our findings and discuss some prospects for future work in \autoref{sec_concl}.

\section{$U(1)$ gauge symmetries upon decoupling gravity in 6d $(1,0)$ theories } \label{sec_U1gaugsymmetries}

We begin by considering a 6d $N=(1,0)$ supergravity theory and analyzing its behaviour as we decouple gravity. As we will see in section \ref{Anomaliesdecoupling}, the structure of anomaly cancellation alone implies that in the decoupling limit any abelian gauge symmetry necessarily becomes a global symmetry with non-zero 't Hooft anomalies, and only non-abelian gauge symmetries can remain dynamical. 
In section \ref{sec_geomintF} we deduce from this field theoretic statement the geometric insight that certain curve classes in an F-theory realisation are necessarily non-contractible.

\subsection{6d $N=(1,0)$ supergravity background}

Let us set the stage by recalling the form of the 
 bosonic part of the six-dimensional $N=(1,0)$ supergravity effective action with $T$ tensor multiplets and gauge group $G=\prod_\kappa G_\kappa\times \prod_{A} U(1)_A$.  In a frame where the six-dimensional Planck mass is set to $M_{\rm Pl} =1$, this effective action takes the form (see e.g. \cite{Bonetti:2011mw} and references therein)
\be \label{6daction1}
\begin{split}
S =& \int_{\mathbb R^{1,5}}   \frac{1}{2} R \ast 1    - \frac{1}{4} g_{\alpha \beta} H^\alpha \wedge \ast H^\beta - \frac{1}{2} \Omega_{\alpha \beta} B^\alpha \wedge X_4^\beta 
  - \frac{1}{2} g_{\alpha \beta} dj^\alpha \wedge \ast dj^\beta \cr
&  -  \sum_\kappa (2 j \cdot b_\ik) \,  \frac{1}{\lambda_\ik}  {\rm tr} F_\kappa \wedge \ast F_\kappa   -  \sum_{A, B} (2 j \cdot b_{AB}) \,  {\rm tr} F^A \wedge \ast F^B + S_{\rm hyper} \,.
\end{split}
\ee
 Here $\alpha =0,1, \ldots, T$ labels the tensor fields $B^\alpha$ with gauge invariant field strength 
 \begin{align}\label{tensors}
H^\alpha = dB^\alpha +\frac{1}{2}a^\alpha\omega_{3L}+2\sum_\kappa\frac{b_\kappa^\alpha}{\lambda_\kappa}\omega_{3Y}^{\kappa}+{2\sum_{A,B}b^\alpha_{AB}\omega_{3Y}^{AB}}.
\end{align}
The field strength is defined in terms of $\omega_{3L}$ and $\omega_{3Y}$ the Chern-Simons forms of the spin connection, the non-abelian gauge fields $A^\kappa$ and the abliean gauge fields $A^A$. The normalization factors $\lambda_\kappa$ are the Dynkin labels of the fundamental representation of each simple gauge group factor. 
Furthermore $a,\, b_{\kappa}$ and $b_{AB}$ are constant $SO(1,T)$ vectors whose indices are contracted by means of the $SO(1,T)$-covariant intersection matrix $\Omega_{\alpha \beta}$ e.g. as 
\bea
a\cdot b= a^{\alpha} b_{\alpha} = a^{\alpha} \, b^{\beta} \, \Omega_{\alpha\beta}.
\eea
The $SO(1,T)$ vector $j^\alpha$  is subject to the constraint
\begin{align}\label{vol1}
j\cdot j=1 \,.
\end{align}
Its independent entries parameterise the dynamical scalar fields in the $T$ tensor multiplets and determine 
the kinetic metric $g_{\alpha \beta}$ as 
\begin{align}
g_{\alpha\beta}=2j_\alpha j_\beta -\Omega_{\alpha\beta} \,.
\end{align}
The dynamical scalar fields $j^\alpha$ furthermore govern the gauge couplings of the non-abelian and of the abelian gauge fields in terms of the constant vectors $b_\kappa$ and $b_{AB}$. 
The expression (\ref{6daction1}) is only a pseudo-action to the extent that the tensor fields are subject to the self-duality constraint
\begin{align} \label{selfdualitycond}
g_{\alpha\beta}*H^\beta = \Omega_{\alpha\beta}H^\beta \,,
\end{align}
which is to be imposed at the level of equations of motion.

As a final piece of information, 
the Chern-Simons couplings of the tensor fields involve the 4-form
\bea \label{X4alpha}
X_4^\alpha = \frac{1}{2}a^\alpha \,  {\rm tr} R \wedge R+2\sum_\kappa\frac{b_\kappa^\alpha}{\lambda_\kappa} \,  {\rm tr} F^{\kappa}  \wedge F^{\kappa} +2\sum_{A,B}b^\alpha_{AB} \, F^{A} \wedge F^B.
\eea
Due to a gauging of the 2-form potential $B^\alpha$ with respect to the gauge symmetries and diffeomorphism this Chern-Simons term renders the classical pseudo-action anomalous in such a way that the classical gauge variance cancels the 1-loop gauge and gravitational anomalies provided the latter factorise as 
\bea \label{I8GSSW}  
I_8^{\rm 1-loop} = \frac{1}{32} \Omega_{\alpha \beta} X_4^\alpha \wedge X_4^\beta \,.
\eea
Here $I_8$ denotes the anomaly polynomial.
This cancellation via the Green-Schwarz-Sagnotti-West mechanism  \cite{Green:1984bx,Sagnotti:1992qw} is encoded in the anomaly equations, whose part involving the abelian gauge fields takes the form 
\begin{align}\label{anomaly1}
a\cdot b_{AB} &= -\frac{1}{6}\sum_I\mathcal M_I q_{IA}q_{IB}\\ \label{anomaly2}
0 &= \sum_I\mathcal M_I^\kappa E^I_\kappa q_{IA}\\ \label{anomaly3}
\frac{b_\kappa}{\lambda_\kappa}\cdot b_{AB} &= \sum_I \mathcal M_I^\kappa A_\kappa^I q_{IA}q_{IB}\\ \label{anomaly4}
b_{AB}\cdot b_{CD}+b_{AC}\cdot b_{BD}+b_{AD}\cdot b_{BC}&= \sum_I\mathcal M_I q_{IA} q_{IB} q_{IC}q_{ID} \ . 
\end{align}
The sums in the above equations are taken over the irreducible representations $I$ of the gauge group, of which $U(1)_A$ charges are denoted by $q_{IA}$. We denote the dimension of $I$ by $\mathcal M_I$ and the number of $G_\ik$ representations in $I$ by $\mathcal M_I^\ik$. Furthermore, $A^I_\kappa$ and $E^I_\kappa$ are defined through
\begin{align}
\tr_I F_\kappa^2 = A_\kappa^I\tr F_\kappa^2\,,\qquad \qquad \tr_I F_\kappa^3 = E^I_\kappa \tr F_\kappa^3 \,,
\end{align}
where $\tr$ denotes the trace in the fundamental representation.

\subsection{Anomalies in the decoupling limit} \label{Anomaliesdecoupling}

After this preparation we can now discuss 
the decoupling of gravity in more detail. To this end it is important to distinguish between the tensor in the gravity multiplet and those in the tensor multiplets. These differ in that the former is self-dual while the latter are anti-self-dual, where the self-duality condition on the tensor fields  $H^\alpha$ is given in (\ref{selfdualitycond}). Let us rewrite this  as
\bea \label{duality2}
\ast H^\alpha = D^\alpha{}_{\beta} H^\beta \,
\eea
in terms of the duality matrix
\begin{align} \label{Dmatrix}
D(j)^\alpha{}_{\beta} := (g^{-1})^{\alpha\gamma}\Omega_{\gamma\beta} = 2j^\alpha j_\beta - \delta^\alpha{}_\beta\, \,.
\end{align}
This matrix has the properties 
\bea
D(j)^\alpha{}_{\beta}  D(j)^\beta{}_{\gamma} = \delta^\alpha{}_{\gamma}, \qquad D(j)^\alpha{}_{\alpha} = 1 - T
\eea
due to \eqref{duality2} and thus has a single ``positive eigenvector'' with eigenvalue $+1$ and $T$ ``negative eigenvectors'' with eigenvalue $-1$; the tensor in the gravity multiplet corresponds to the positive eigenvector. Using again \eqref{duality2} we immediately see that such a positive eigenvector is $j$, that is, 
\bea
D(j)^\alpha{}_{\beta} j^\beta= + j^\alpha \,.
\eea
This vector hence spans the one-dimensional positive eigenspace while the $T$-dimensional space orthogonal to $j$ with respect to $\Omega_{\alpha\beta}$ is spanned by the negative eigenvectors.
The latter are associated to the 2-form potentials in the $T$ tensor multiplets and contain the tensors different from the gravi-tensor. Given this split of the vector space $\mathbb R^{1,T}$, every vector can be decomposed into its positive and negative parts, which we will denote by the superscripts $+$ and $-$, respectively, i.e.
 \bea \label{vpmsplit}
 v = v^+(j) + v^-(j), \qquad D(j)^\alpha{}_{\beta} \, v^\pm(j)^\beta = \pm v^\pm(j)^\alpha \,.
 \eea
Importantly, since the duality matrix $D(j)^\alpha{}_{\beta}$ depends on the choice of $j$, so does this split. 
 
Given the split (\ref{vpmsplit}), the Bianchi identities for the tensors~\eqref{tensors} decompose, for each choice of $j$, as
\begin{align}\label{Bsplit}
dH^{\pm}(j) = \frac{1}{2}a^{\pm}(j) \, \tr R^2 + \sum_\kappa\frac{2 b_{\kappa}^\pm(j)}{\lambda_\kappa}\tr F^2_\kappa + {\sum_{A,B}2b_{AB}^{\pm}(j)F^A\wedge F^B} \,.
\end{align}
In particular, $H^{+}$ is the field strength of the tensor in the gravity multiplet.

The decoupling of gravity corresponds to taking the gauge coupling constant of certain gauge group factors $G_{\hat\kappa}$ to infinity while keeping the six-dimensional Planck mass finite.
This happens in suitable regions of the tensor moduli space controlled by the values of the scalar fields $j^\alpha$. Let us denote by $j^\alpha_0$ a choice of scalar fields realising this limit where $j_0\cdot b_{\hat\kappa} =0$. At this point in moduli space, $b_{\hat \kappa}$ becomes a \emph{negative} eigenvector of the duality matrix, namely
\begin{align}\label{negeigen}
j_0\cdot b_{\hat\kappa}=0\qquad \Longrightarrow \qquad D(j_0)b_{\hat\kappa} = -b_{\hat\kappa}.
\end{align}
Thus, by (\ref{Bsplit}) such a gauge theory decouples from the gravi-tensor $B^{+}(j_0)$. This means, in particular, that $B^+(j_0)$ does not participate in the Green-Schwarz mechanism cancelling anomalies which involve $G_{\hat\kappa}$-gauge fields. Indeed, such a contribution is proportional to 
\begin{align}
b_{\hat\kappa} \cdot v = b_{{\hat\kappa}}^{+}(j)\cdot v^+(j) + b_{{\hat\kappa}}^{-}(j)\cdot v^-(j)
\end{align}
for some vector $v$, where the first term is the contribution due to the gravi-tensor while the second is due to the tensor multiplets. Once we decouple gravity, $b_{{\hat\kappa}}^{+}(j_0)=0$ so the gravi-tensor does not contribute. This guarantees that there are no anomalies involving the gauge group $G_{\hat\kappa}$ after decoupling gravity.

More generally, however, once we decouple gravity some of the gauge symmetries may become global symmetries and acquire a non-zero 't Hooft anomaly. It is particularly interesting to analyse the case of a $U(1)$ gauge symmetry since it allows us to conclude that such $U(1)$ symmetries can only survive as global symmetries after decoupling gravity.  To see this, consider the quartic abelian $U(1)_A$ anomaly 
\bea \label{U1Athooft}
b_{A }^{+}(j) \cdot b_{A}^{+}(j)  +  b_{A}^{-}(j) \cdot b_{A}^{-}(j)  = \frac {1}{3} \sum_I\mathcal M_I q_{IA}^4  \geq 0 \,,
\eea
with 
\bea
b_A\equiv b_{AA}.
\eea
According to the above logic, in the decoupling limit $j \rightarrow j_0$ we ${\it discard}$, on the LHS, the  contribution $b_{A}^{+}(j_0) \cdot b_{A}^{+}(j_0) \geq 0$ and are left only with the term $b_{A}^{-}(j_0) \cdot b_{A}^{-}(j_0)  \leq 0$.
 If at least one hypermultiplet carries non-trivial $U(1)_A$ charge, the RHS is positive. In this case the 1-loop $U(1)_A$ anomaly is no longer cancelled in the decoupling limit, which is possible only if $U(1)_A$ reduces to (at best) a global, or flavour, symmetry. In this case, the violation of (\ref{U1Athooft}) merely indicates a non-zero 't Hooft anomaly.
On the other hand, if no massless hypermultiplets are charged under $U(1)_A$ and the RHS therefore vanishes, the anomaly equation is still satisfied in the decoupling limit provided $b_{A}^{-}(j_0) \cdot b_{A}^{-}(j_0)  = 0$.  Since the metric is negative definite on the negative subspace this is possible only if $b_{A}^{-}(j_0) = 0$ and therefore $b_A = b_{A}^{+}(j_0) + b_{A}^{-}(j_0) = b_{A}^{+}(j_0)$. But 
  the anomaly equation $b_A \cdot b_A  = 0$ away from the decoupling limit 
 then implies that $b_A \equiv 0$.    
If this is the only $U(1)$ in the theory, the theory is completely trivial. In the presence of another $U(1)_B$, the anomaly equations imply that $b_{AB} \cdot b_{AB} = 0$ which, together with the fact that the kinetic term must be positive semidefinite, shows that $b_{AB}=0$ and so $U(1)_A$ is not part of the theory. Indeed, the kinetic matrix for $U(1)_A$ and $U(1)_B$ is 
\bea
\left (\begin{array}{cc} 0&j\cdot b_{AB}\\ j\cdot b_{AB}& j\cdot b_{BB}\end{array}\right ) 
\eea
and in order for it to be positive semidefinite we must have that $j\cdot b_{AB} = 0$ or, equivalently, $b_{AB}^+=0$. This then, together with $b_{AB}^2=0$, implies that $b_{AB}=0$. Notice that this argument works for any number of $U(1)$s that may mix with $U(1)_A$ since positive semidefiniteness of the full kinetic matrix requires its leading principal minors to be all non-negative.

In conclusion, abelian gauge symmetries necessarily become at best global symmetries after decoupling gravity.
One might wonder if in the decoupling limit such a $U(1)_A$ symmetry acquires non-zero chiral (i.e. ABJ) anomalies, namely, mixed anomalies involving any of the gauge groups $G_{\hat\kappa}$ which remain dynamical in the decoupling limit. As we have shown above, however, since $b_{{\hat\kappa}}^{+}(j_0)=0$, there can be no anomalies involving any $G_{\hat\kappa}$ gauge field, including any mixed anomaly with a symmetry that became global after decoupling gravity. 
Hence the mixed $U(1)_A - G_{\hat\kappa}$ anomaly is consistently cancelled even in the decoupling limit, given that it was cancelled in the supergravity regime. 
This is crucial because otherwise the global $U(1)_A$ symmetry would be broken by such a mixed anomaly with a gauge theory factor. 
By contrast, we have shown that the $U(1)_A$ symmetry survives as an exact global or flavour symmetry in the decoupling limit which has only 't Hooft anomalies.

\subsection{Geometric interpretation of the decoupling limit via F-theory} \label{sec_geomintF}

In the sequel we will engineer the 6d $N=(1,0)$ supergravity theory as an F-theory compactification with base space $B_2$.
The F-theoretic realisation of such supergravities has been studied intensively in the literature (see e.g. the more recent \cite{Kumar:2009ac,Kumar:2010ru,Bonetti:2011mw,Park:2011wv} and the review \cite{Taylor:2011wt} for further references).
The $T+1$ tensor fields arise by expanding the Type IIB Ramond-Ramond form $C_4 = B^\alpha \wedge \omega_\alpha$ in terms of a basis 
$\omega_\alpha$ of $H^{1,1}(B_2)$, and hence 
$T+1 = h^{1,1}(B_2)$.
The intersection matrix $\Omega_{\alpha \beta}$ is given by the topological intersection pairing
\beq
\Omega_{\alpha \beta}=\int_{B_2} \omega_\alpha \wedge \omega_\beta \  
\eeq 
of the base surface,
and also the $SO(1,T)$  vectors $a,\,b_\kappa$ and $b_{A}$  appearing for instance in (\ref{tensors}) have a geometric interpretation: the object $a$ corresponds to the canonical class of $B_2$,
\bea
K = a^\alpha  \omega_\alpha \,,
\eea
 and $b_\kappa = b^\alpha _\kappa  \, \omega_\alpha$ is the class of the curve $C_\kappa$ in the base 
wrapped by a stack of 7-branes  
 giving rise to a gauge group $G_\kappa$. As we will review in the next section, a similar interpretation as an effective  divisor class on $B_2$ exists  for the anomaly coefficient   $b_{A} \equiv b_{AA}$ of a $U(1)_A$ gauge symmetry.
 Furthermore, the scalar fields in the tensor multiplets parametrise the 
K\"ahler form 
\bea
J=j^\alpha \omega_\alpha
\eea
 of the F-theory base,
where the constraint (\ref{vol1}) implies that we have set the volume of the base to $1$.\footnote{Note that the overall volume of the base is part of the universal hypermultiplet.} This is the geometric analogue of the statement that we are working at a fixed value of the 6d Planck mass
\beq\label{decouple}
M_{\rm Pl}^{-4} \propto {\rm Vol}_J (B_2) \, .
\eeq
The gauge kinetic functions of the non-abelian and abelian gauge group factors are then determined by the volumes of the respective curve classes,
\beq \label{gAgkappa}
f_{AA} \propto  {\rm Vol}_J (b_A)=  j \cdot b_A  \,,  \qquad {g_\kappa^{-2}} \propto  {\rm Vol}_J (b_\kappa)  =  j\cdot b_\kappa\,,
\eeq
where $f_{AA}$ multiplies the diagonal part of the abelian gauge kinetic term of the Lagrangian.
We will review the derivation of this classic relation in  \autoref{reviewMW}.

 The decoupling of gravity with some of the non-abelian gauge group factors $G_{\hat \kappa}$ kept dynamical can then be described geometrically in two equivalent ways.
If we choose to work in the above frame where the 6d Planck mass is fixed at $M_{\rm Pl} =1$, then in the decoupling limit the K\"ahler form $J$ approaches a boundary of the K\"ahler cone along which the volume of the curves $C_{\hat\kappa}$ tends to zero, while the volume of the base $B_2$ stays finite.
In this limit the
 K\"ahler form $J$ of $B_2$  approaches a value $J_0$ on the boundary of the K\"ahler cone for which\footnote{The exponent, $\frac{1}{2}$, in the denominator is chosen so that the numerator and the denominator have the same mass dimensions. Note that for any other exponent choices, there always exists an appropriate rescaling of the K\"ahler form $J$ such that the ratio becomes zero.}  
\beq \label{limit-def}
{\rm lim}_{J\to J_0} \frac{
{\rm Vol}_{J}(b_{\hat\kappa})
}{({\rm Vol}_J (B_2))^{\frac{1}{2}}}  =  0 \,.
\eeq
The curves for which this is possible are called contractible.

Alternatively, we can think of the decoupling limit in more physical terms by taking ${\rm Vol}_J(B_2) \to \infty$ while keeping  ${\rm Vol}_{J}(b_{\hat\kappa})$ finite.
In this frame, the 6d Planck mass tends to infinity while the gauge coupling of $G_{\hat\kappa}$ remains finite, as is more appropriate from a physical perspective.
In particular, in this picture we can consider a two-step limit, which is relevant for the definition of superconformal theories: First decouple gravity from the gauge theory $G_{\hat\kappa}$ while keeping ${\rm Vol}_{J}(b_{\hat\kappa})$ finite, and in the second step take ${\rm Vol}_{J}(b_{\hat\kappa}) \rightarrow 0$, corresponding to the strong coupling SCFT limit of the latter. 
Both descriptions satisfy (\ref{limit-def}) and are in fact mathematically equivalent (see e.g. \cite{Cordova:2009fg}).

In the absence of kinetic mixing, a similar interpretation can be given for abelian gauge symmetries: In this case $f_{AA}$ in (\ref{gAgkappa}) is the inverse squared $U(1)_A$ gauge coupling.   
Since the $U(1)_A$ symmetry becomes a global symmetry as we decouple gravity, the curve $b_A$ remains of finite volume in every possible limit in the K\"ahler moduli space keeping the overall base volume finite. It is therefore non-contractible. 
The same conclusion can be reached in an even simpler manner again by exploiting the anomaly condition (\ref{anomaly4}) for 
 $U(1)_A$ (in absence of kinetic mixing): 
The LHS is the self-intersection of the curve class $b_A$ on $B_2$, and the anomaly equation implies that this self-intersection is non-negative. But on a complex base $B_2$, contractible curves necessarily have negative self-intersection, as will be discussed in more detail in \autoref{sec_DecoupFCY3}.

In the presence of kinetic mixing a simple geometric interpretation of the gauge coupling is more elusive. The reason is that while the kinetic matrix $f_{AB}$ can be diagonalized for each fixed choice of $j$, this involves irrational coefficients which render an interpretation in terms of (integral) divisors on the base less clear.

Let us summarize our discussion so far.
Of the collection of curve classes $b_\kappa$, suppose that a subcollection, $b_{\hat \kappa}$, simultaneously contract to zero volume for a chosen $J_0 = j_0^\alpha \omega_\alpha$ on the boundary of K\"ahler cone, while the rest remain to have a finite volume. 
Then the theory associated to this geometry has gravity decoupled and gauge group $\prod_{\hat \kappa} G_{\hat \kappa}$. On the other hand, the abelian group $\prod_A U(1)_A$, as well as the remaining part of the original non-abelian group, survive as a global symmetry with 't Hooft anomalies only. 
In the absence of kinetic mixing,  each abelian gauge coupling is determined  by  the volume of the respective curve $b_A$, which remains finite in the limit.
In order to compute the 't Hooft anomalies, we must only include the contribution of the tensors corresponding to the curves that shrink in the Green-Schwarz mechanism. This is done by simply projecting any of the  $SO(1,T)$ vectors appearing in the LHS of the anomaly equation onto the vector space spanned by the tensors which remain dynamical. In \autoref{sec_examples} we will consider some examples where this abstract discussion is made explicit.

\section{Decoupling of U(1)s in F-theory on elliptic Calabi-Yau 3-folds} \label{sec_DecoupFCY3}

In this section we prove that every curve $b_{AA} \equiv b_A$ associated with a $U(1)_A$ symmetry is non-contractible directly by analyzing the elliptic fibration over $B_2$ underlying the F-theory interpretation, and without any reference to the anomaly equations. The proof holds irrespective of the number abelian gauge group factors. In absence of kinetic mixing, the volume of $b_A$ has the interpretation of the inverse squared gauge coupling, and our result therefore proves that this $U(1)_A$ cannot remain as a gauge symmetry in the decoupling limit. A special case without kinetic mixing is a setup with only a single $U(1)_A$. 

The fact that the curve associated with a $U(1)$ gauge symmetry is necessarily non-contractible is a rather non-trivial statement if we are to approach the question from the perspective of a perturbative Type II brane setup. The abelian gauge group is associated with a linear combination of curves, each of which carries a gauge group $U(N)$ such that the sum of the diagonal $U(1)$ factors survives the geometric St\"uckelberg mechanism and remains massless. Our results show that such a linear combination of curves can never be shrinkable. This is a priori not completely obvious given that, for instance, shrinkable curves of self-intersection (-1) in F-theory can well carry non-abelian gauge groups $SU(N)$; hence one might believe at first sight that they could team up, in the perturbative limit, to form a $U(1)$ gauge group as sketched above.
The power of F-theory is to translate this question directly into the property of a well-defined geometric object, the height pairing, which can be studied systematically.

For pedagogical reasons we first discuss the non-decoupling of abelian gauge theories in six-dimensional theories
which do not exhibit any non-abelian gauge group factors in \autoref{subsec_WeakFanoDecoup}. As we will see, the absence of non-abelian gauge symmetry leads to a few technical simplifications.
The general six-dimensional case is then treated in \autoref{6g_general}. For the reader's convenience we review, in \autoref{reviewMW}, the relation between $U(1)_A$ gauge symmetries and the geometry of rational sections in F-theory. 

\subsection{U(1)s on a generalized del Pezzo base} \label{subsec_WeakFanoDecoup}

Let us therefore consider F-theory compactified on a Calabi-Yau $3$-fold $\hat Y_3$ which is elliptically fibered over a complex $2$-fold base $B_2$ with projection
\beq \label{ellfibration1}
\pi: \hat Y_{3} \to B_2 \,.
\eeq
The effective action of this compactification is an ${\cal N}=(1,0)$ supergravity theory  in $\mathbb R^{1,5}$  whose gauge group\footnote{We are only interested in the part of the gauge group from the 7-brane sector as these are the only ones under which light matter can be charged. In addition, there can be abelian gauge group factors from what in Type IIB language would be called the closed string or Ramond-Ramond sector, whose charged states are wrapped branes which are always heavy.} we 
assume, for now, to be of the form
\beq
G= \prod_{A=1}^r U(1)_A \ . 
\eeq
The abelian gauge factors arise from the non-torsional rational sections of the fibration \cite{Morrison:1996pp}. The reader not familiar with this concept is advised to jump now to 
 \autoref{reviewMW} for some background and a self-contained derivation of the following facts:
Given a section $s_A$ and its associated divisor $S_A = {\rm div}(s_A)$, one first defines the Shioda homomorphism
\beq\label{shioda1a}
\sigma(s_A) = S_A-Z-\pi^{-1} \left(\pi_\ast((S_A-Z)\cdot Z)\right)  \in H_4(\hat Y_3) \,,
\eeq
where $Z$ denotes the zero-section of the elliptic fibration and the pushforward $\pi_\ast$ is defined in (\ref{pushforwardmap}).
In the dual  M-theory expanding the $3$-form potential as
$C_3 = \sum_{A=1}^r A^A \wedge [\sigma(s_A)] + \ldots$
gives rise to abelian gauge potentials $A^A$ which lift to the gauge potentials of the gauge group factor $U(1)_A$ in F-theory.
The gauge kinetic terms 
of the abelian gauge factors in F-theory are given by
\bea \label{SkinF}
S_{\rm kin}  =     - \frac{2\pi}{2}  \hat f_{AB}  \int_{\mathbb R^{1,5}} dA^A \wedge \ast d A^B  \,, \qquad \hat f_{AB}    =   \int_{B_2} J \wedge b_{AB}  \,,
\eea
with $J$ the K\"ahler form on the base $B_2$. The object  
\bea 
b_{AB} = - \pi_\ast (\sigma(s_A) \cdot   \sigma(s_B))
\eea
 is known in arithmetic geometry as the \emph{height pairing} of the section $s_A$ with $s_B$ and defines a curve class on $B_2$. 
 Of special interest for us is the height pairing of $s_A$ with itself,
 \bea \label{UABdef1}
 b_A := b_{AA} =  - \pi_\ast (\sigma(s_A) \cdot   \sigma(s_A)) \,.
 \eea
As reviewed in  Appendix \ref{app_Gaugecoupl}, the height pairing can be evaluated as  
\beq \label{hp1}
\Ub_A=2\bar K + 2\pi_\ast(S_A \cdot Z) \,,
\eeq
where ${\bar K} = - K$ is the anti-canonical divisor of $B_2$ and $\pi_\ast(S_A \cdot Z) $ is the curve on $B_2$ over which the section $S_A$ and the zero-section $Z$ meet in the fiber.

In the absence of kinetic mixing, the volume of the curve $b_A$ with respect to the K\"ahler form $J$ on $B_2$ has the simple interpretation of determining the $U(1)_A$ gauge coupling 
\beq \label{gA2a}
{g_A^{-2}} \propto {\rm Vol}_J (b_A) \,.
 \eeq

Even though our geometric results for $b_A$ hold irrespective of the presence of kinetic mixing, we shall henceforth restrict to this case.
The decoupling criterion (\ref{limit-def}) then turns into the contractibility criterion for each divisor $\Ub_A$ described by the height pairing (\ref{UABdef1}) and (\ref{hp1}). 
Since we are only interested in those base manifolds $B_2$ over which an elliptically fibered Calabi-Yau 3-fold exists, $\bar K$ has to be an effective divisor. 
This is because every elliptic fibration is birationally equivalent to a Weierstrass model
\bea \label{Weier}
y^2 = x^3 + f \, x \, z^4 + g \, z^6
\eea
with $[x : y : z]$ homogeneous coordinates on the fiber ambient space $\mathbb P_{231}^2$ and $f \in H^0(B_2, {\cal O}(4 \bar K))$ and $g \in H^0(B_2, {\cal O}(6 \bar K))$.
Hence, in order for $f$ and $g$ to exist as holomorphic sections of the indicated line bundles ${\cal O}(4 \bar K)$ and  ${\cal O}(4 \bar K)$, the divisor $\bar K$ must be effective.
Moreover, the second term in the height pairing~\eqref{hp1} is the divisor in the base over which the two effective divisors $S_A$ and $Z$ of $\hat Y_3$ intersect.
This locus defines an algebraic curve, hence the divisor $\pi_\ast(S_A \cdot Z)$ is also effective. We thus learn that the height pairing $\Ub_A$ is effective. 
Of course, in view of (\ref{gA2a}) this is required for consistency of the effective action because $g_A^{-2}$ must be strictly positive everywhere in the interior of the K\"ahler cone of $B_2$.

To study the contractibility of $\Ub_A$ we recall the following fact:
Consider an effective divisor 
\be
C = \sum_i c_i \, C_i \,,  \qquad c_i \in \IZ_{\geq 0} \,,
\ee
on a complex surface $S$ with irreducible components $C_i$, each describing an effective curve;
then $C$ is contractible if and only if the union of $C_i$ is contractible to one or several points.
According to Mumford's contractibility criterion \cite{mumford},
a necessary condition for the union of irreducible curves $C_i$ to contract to possibly several points is 
that the intersection matrix of the curves  be negative semi-definite:
\be
\{ C_i \}  \quad {\rm contractible}    \quad \Rightarrow  \quad  I_{ij} := C_i \cdot C_j    \stackrel{!}{\leq} 0 \,.
 \ee
In order for the union of irreducible curves $C_i$ to be simultaneously contractible to a single point, a necessary condition is  $I_{ij}$ to be negative-definite.

To show that this criterion is violated by the height pairing (\ref{hp1}), recall that in this section we are assuming that the gauge group in F-theory is purely abelian.
This assumption has far-reaching consequences for the base $B_2$: As studied in \cite{Morrison:2012np}, if the base $B_2$ contains an irreducible  curve $C$ of self-intersection $C \cdot C = - n$ with $n \geq 3$, then the fibers of the Weierstrass model $Y_3$ associated to $\hat Y_3$ must necessarily degenerate over $C$ in such a way that $Y_3$ is singular and $C$ carries a non-abelian gauge group. This phenomenon has been dubbed `non-Higgsable cluster' (NHC) because it is the geometric manifestation of the fact that the non-abelian gauge symmetry from the minimal 7-brane stack wrapping $C$ does not allow for a Higgs branch along which the gauge group could be broken. Interestingly, the only surfaces that do not have at least one curve of self-intersection $C \cdot C = -3$ or below are the generalized del Pezzo  or weak Fano surfacees, i.e. non-singular projective surfaces with 
\bea \label{weakFano}
\bar K \cdot \bar K > 0,  \qquad \bar K \cdot C \geq 0
\eea
for every effective curve $C$ \cite{Morrison:2014lca}. Our assumption of absence of non-abelian gauge group factors forces $B_2$ to have only irreducible curves of self-intersection $C \cdot C \geq -2$ 
and hence to be a generalized del Pezzo surface.

Then, given the decomposition of the anti-canonical divisor 
\beq
\bar K = \sum_i \gamma_i \, \Sigma_i \ , \quad \gamma_i \in \IZ_{\geq 0} \ , 
\eeq
into its irreducible components $\Sigma_i$, the intersection matrix $I_{ij}:=\Sigma_i \cdot \Sigma_j$ is not negative semi-definite as 
\be
\sum_{i,j} \gamma_i \, I_{ij} \, \gamma_j = \bar K \cdot \bar K >0
\ee
 and hence $\bar K$ is not contractible. 
Thus $\Ub_A$ is the sum of two effective divisors as in  (\ref{hp1}), at least one of which is not contractible. This shows that $b_A$ is not contractible.

\subsection{Including non-abelian gauge groups in 6d} \label{6g_general}

Let us now extend our analysis to a more general F-theory compactification to six dimensions, with gauge group
\bea
G =   \prod_{A=1}^r U(1)_A \times \prod_\ik G_\ik \,,
\eea
where each $G_\ik$ represents a non-abelian simple Lie group.
This leads to two types of modifications: First, as we recall momentarily, the Shioda homomorphism and hence the height pairing acquires additional contributions which at first sight complicate the analysis. Second, the weak Fano condition (\ref{weakFano}) can no longer be assumed and in particular $\bar K \cdot \bar K $ can be, and in general is, negative. 
Nonetheless, we will show that $\bar K$ cannot be contractible and deduce from this that $\Ub_A$ has the same property. Again, in the case of only a single $U(1)_A$ this implies that $U(1)_A$ cannot survive as a gauge theory after decoupling gravity, but the result on $b_A$ as such holds in general.

Let us first recall the well-known modifications of the Shioda homomorphism in the presence of non-abelian gauge symmetry.
 The non-abelian gauge group factors are due to stacks of in general mutually non-local $[p,q]$ 7-branes, each wrapping an irreducible component $W_I$ of the discriminant divisor of the elliptic fibration. 
In the presence of such non-abelian gauge groups, one distinguishes between the singular Weierstrass model $Y_3$, (\ref{Weier}), and its resolution $\hat Y_3$.
 The vanishing locus of the discriminant polynomial
\bea
\Delta = 4 f^3 + 27 g^2 \,
\eea
associated with the Weierstrass model 
 describes a divisor $\Sigma$ on $B_2$, with irreducible components $C_\ik$.
The Weierstrass model is singular in the fiber over $C_\ik$ and has a Calabi-Yau resolution $\hat Y_3$.
The singular point in the fiber over $C_\ik$ is replaced by a chain of rational curves, which form, together with the original fiber, the affine Dynkin diagram of the Lie algebra $\mathfrak{g}_\ik$ of $G_\ik$.  
More precisely, resolving the singularities in the fiber of $C_\ik$ introduces the resolution or 
Cartan divisors 
\bea \label{Eik}
E_{i_\ik},  \qquad i_\ik = 1, \ldots, {\rm rk}(\mathfrak{g}_\ik),
\eea
on $\hat Y_3$ which are generically $\mathbb P^1$-fibrations over $C_\ik$ with fiber $\mathbb P^1_{i_\ik}$.
The pullback of $C_\ik$ is of the form
\bea
\pi^{-1} (C_\ik) = \sum_{i_\ik=0}^{ {\rm rk}(\mathfrak{g}_\ik)}  a_{i_\ik}  \, E_{i_\ik}
\eea
with $a_{i_\ik}$ the comarks of the affine Dynkin diagram (and $a_{0_\ik}=1$). The divisor $E_{0_\ik}$ is likewise generically rationally fibered with fiber $\mathbb P^1_{0_\ik}$. This rational curve is distinguished by the fact that it is intersected by the zero-section divisor $Z$.

We will review in Appendix \ref{Correctionsnonab} that, in order to give rise to a properly normalised $U(1)_A$ gauge group factor, the Shioda map $\sigma(s_A)$ is required to satisfy the extra condition 
\bea \label{DP1=0a}
\sigma(s_A) \cdot   \mathbb P^1_{i_\ik} = 0 \,,   \,\, \,   i_\ik = 1, \ldots, {\rm rk}(\mathfrak{g}_\ik) \,.
\eea
This can always be achieved by modifying the expression (\ref{shioda1a})  into 
\beq\label{shioda-A}
\sigma(s_A) = S_A-Z-\pi^{-1} (\pi_\ast((S_A-Z)\cdot Z))+ \sum_\ik \sum_{i_\ik}  \ell_{A}^{i_\ik}   E_{i_\ik} \,, 
\eeq 
where the coefficients $\ell_{A}^{i_\ik} \in \mathbb Q$ can be found in (\ref{lAjik}).

Now, if we compute the height pairing for (\ref{shioda-A}), the correction terms will lead to an expression involving effective divisor classes, but in general with positive and negative coefficients. 
While the total class of the height pairing is still effective, this has a serious drawback for us: Given an effective divisor class $\delta$ presented, say, as the {\it difference} of two effective divisor classes, $\delta = \alpha - \beta$, we cannot show that the class  $\delta$ is non-contractible by showing that one of the two classes $\alpha$ or $\beta$ is non-contractible. This is possible only if we have a decomposition  for a {\it sum} of two effective divisors. 

The way out is the observation that we may still find an alternative expression for the Shioda homomorphism without any contributions from $E_{j_\ik}$. 
This expression can be obtained by making use of the following theorem: For each discriminant component $C_\ik$ and gauge group factor $G_\ik$, there exists a finite integer $m_\ik$ for which the image point of the section 
 $m_\ik\, s_A$, i.e. the multiple of $s_A$ in $MW(\pi)$,
 lies in the affine component $\mathbb P^1_{0_\ik}$ of the generic fiber over $C_\ik$. 
 Put differently, the divisor associated with $m_\ik s_A$ has the intersection numbers
 \bea \label{div-intnumbers1}
 {\rm div}(m_\ik \,  s_A) \cdot \IP^1_{0_\ik} = 1, \qquad  {\rm div}(m_\ik \,  s_A) \cdot \IP^1_{i_\ik} = 0 \quad \forall \,  i_\ik \in  \{1, \ldots, {\rm rk}(\mathfrak{g}_\ik)\} \,.
 \eea
 Mathematically this is a consequence of the following beautiful fact in arithmetic geometry, proven by Kodaira \cite{Kodaira} and N\'eron \cite{Neron} for elliptically fibered surfaces:
 The notion of addition of points on the general elliptic fiber can be extended to the degenerate, reducible  fibers of $\hat Y_3$ over $C_\ik$. The set of sections lying in the affine component of the degenerate fiber over $C_\ik$ form a subgroup $MW(\pi)_{0,\ik}$ of the Mordell-Weil group, and $MW(\pi) / MW(\pi)_{0,\ik}$ is a group of finite order $m_\ik$.
 In extending this statement to elliptically fibered varieties of higher dimension, only two changes occur: First, due to monodromies along $C_\ik$ the global structure of the fiber may change compared to the local generic fiber over $C_\ik$. 
 In this case the gauge algebra is a non-simply laced subalgebra $\tilde{\mathfrak{g}}_\ik$ of the algebra $\mathfrak{g}_\ik$ associated with the local fiber, and the relevant order satisfies  $\tilde m_\ik \leq m_\ik$. 
 Second, the fiber type changes in codimension-one and more along $C_\ik$. This, however, is of no relevance for the behavior of the section over \emph{generic} points of $C_\ik$ and hence for the intersection numbers (\ref{div-intnumbers1}).
 
 As a result, for a given elliptic fibration $\hat Y_3$ there exists a finite integer $m$ such that the section $m \, s_A$ lies in the affine component of the generic fiber over \emph{every} discriminant component $C_\ik$, or
  \bea \label{div-intnumbers}
\forall \, \ik: \quad  {\rm div}(m \,  s_A) \cdot \IP^1_{0_\ik} = 1, \qquad  {\rm div}(m\,  s_A) \cdot \IP^1_{i_\ik} = 0 \quad \forall \,  i_\ik \in  \{1, \ldots, {\rm rk}(\mathfrak{g}_\ik)\} \,.
 \eea
Using that the Shioda map is a homorphism, this implies that 
\bea
m \,\sigma(s_A) &=& \sigma(m \, s_A) \\ 
&=& {\rm div}(m\, s_A) - Z - \pi^{-1} (\pi_\ast (( {\rm div}(m\,  s_A) - Z) \cdot Z)) \,
\eea
without the necessity of extra correction terms to implement the analogue of (\ref{DP1=0a}). 
The height pairing $\Ub_A $ can then be expressed as
\beq\label{shiodamaprefined}
\Ub_A= -\pi_\ast\left( \sigma(s_A) \cdot \sigma(s_A) \right) = \frac{1}{m^2} (2\bar K + 2\pi_\ast ({\rm div}(m\,s_A) \cdot Z)) \ . 
\eeq

The expression in brackets is still a {\it sum} of two effective divisor classes,\footnote{
Recall that $\bar K$ has to be effective for there to exist an F-theory model over the base $B_2$. Furthermore, the pushforward of the intersection of two sections is also effective as long as the two sections are distinct. The only exception to the latter is when $m s_A = 0$, in which case ${\rm div}(ms_A) \cdot Z = - \bar K$, leading to $b_A=0$. However, in this case, $s_A$ is a torsional section and does not give rise to an abelian gauge group in the first place~\cite{Aspinwall:1998xj,Mayrhofer:2014opa,Cvetic:2017epq}.
} for the same reasons as in section \ref{subsec_WeakFanoDecoup}.
Hence we can again show that $\Ub_A$ is not contractible by showing that $\bar K$ is not contractible. 
A second complication compared to the procedure of the previous section occurs because in the presence of non-abelian gauge group factors, the base surface is not necessarily of the generalized del Pezzo type; therefore we can no longer use that $\bar K \cdot \bar K >0$ to show non-contractibility of ${\bar K}$. In the context of F-theory vacua, however, we can still prove that $\bar K$ is not contractible as follows. 
First, we will argue that if $\bar K$ were contractible at finite distance in moduli space, it would, loosely speaking, not contain any 1-cycles. And then we will see that this is at odds with its property of being the anti-canonical divisor of a surface.

In general, $\bar K$ is a reducible divisor.  Consider therefore a curve $C = \sum_i \gamma_i \, C_i$ with irreducible curve components $C_i$.
The requirement that a surface $B_2$ serves as the base of an elliptic Calabi-Yau fibration $Y_3$
severely restricts the  type of contractible curves $C$, as classified in \cite{Heckman:2013pva}. The classification proceeds in two steps:
Suppose $C$ is contractible. 
If $C$ contains some curve components of self-intersection $C_k \cdot C_k = -1$, called $(-1)$-curves, contract these to points. 
If we denote the surface obtained by contraction of the $(-1)$-curve components of $C$ by $B_{2,1}$, then the contraction defines a map
\bea \label{rhomap}
\rho: B_2 \rightarrow B_{2,1} \,.
\eea
A $(-1)$-curve on a complex surface has the special property that its contraction does not lead to any singularities; hence $B_{2,1}$ is smooth. 
The remaining set of curves is called in \cite{Heckman:2013pva} `endpoint' configuration $C_{\rm end} \subset B_{2,1}$,
\bea
 \rho^{-1}(C_{\rm end}) = C \,.
\eea 
The main theorem is then that contraction of $C_{\rm end}$ on $B_{2,1}$ to a point $p$, in a manner compatible with the existence of a Calabi-Yau Weierstrass model $Y_3$ over $B_2$, leads to a singularity of the local form $\mathbb C^2/\Gamma$
with $\Gamma$ a discrete subgroup of $U(2)$. 
In other words the contraction of $C_{\rm end}$ to a point $p$ defines a map
\bea \label{psi-res}
\psi: B_{2,1} \rightarrow B_{2,2}, \qquad \psi^{-1}(p) = C_{\rm end} \,,
\eea
such that there exists a local neighbourhood $U_p$ of $p$ of the form
\bea \label{Up}
U_p \simeq \mathbb C^2/\Gamma, \qquad \Gamma \subset U(2) \,.
\eea

The restriction to an orbifold of this type is due to the fact that the original surface $B_2$ is required, by assumption, to support an elliptic fibration $Y_3 \rightarrow B_2$. This implies in particular that none of the curves in the contractible set $C$ must be  of self-intersection $C_l \cdot C_l < -12$ as otherwise the sections $f$ and $g$ defining the associated Weierstrass model would vanish to order $ \geq 4$ and $\geq 6$. Inspection of all possible contractible curve configurations $C_{\rm end}$ compatible with the existence of a Weierstrass model then shows that upon contraction they give rise to a singularity of the form (\ref{Up}) \cite{Heckman:2013pva}.

This has the following consequences for us: 
An orbifold singularity of a surface of the above type is an \emph{at worst canonical singularity} (see e.g. \cite{Ishi}).
Recall that given a resolution 
\bea \label{resf}
f: \hat X \rightarrow X
\eea
of a singularity at $p \in X$ one defines the exceptional set as the locus on $\hat X$ along which $X$ and $f^{-1}(X)$ differ. If we denote by $E_i$ the strata of the codimension-one exceptional set, the canonical bundles of both spaces compare as  $K_{\hat X}  = K_X + \sum_i a_i E_i$ with $a_i$ the discrepancies. The singularity is called at worst canonical if $a_i \geq 0$ for all $i$, and at worst terminal if $a_i>0$ for all $i$. 
Now, canonical (and in particular terminal) singularities have the property of being \emph{rational}.
 Intuitively, this means that the exceptional locus in the blow-up of the singularity carries no cohomologically non-trivial $(i,0)$-forms for $i >0$.
This is formalized by stating that given a rational singularity of a variety $X$ at a point $p$ with resolution (\ref{resf}),
the so-called right-derived functor sheaf vanishes
\beq\label{rdfa}
R^i f_\ast \cO_{\hat X} = 0, \quad i>0\,, \ 
\eeq
in a neighborhood $U_p$ of $p$. Here, $R^i f_\ast \cO_{\hat X}$ is locally represented by the pre-sheaf on $X$ that associates to an open set $U_p$ the cohomology group $H^i(f^{-1}(U_p), {\cO})$. In particular the stalk at the singular point $p$ on $X$ is given by $H^i(f^{-1}(p), {\cO})$. 
The locus $f^{-1}(p)$ is in turn precisely the exceptional locus of the resolution $\hat X$, whose blow-down gives rise to the singularity on $X$.

Applied to the resolution (\ref{psi-res}), this shows that 
\bea
H^i(C_{\rm end}, {\cal O}) = 0, \quad i >0.
\eea
Ultimately, we are not interested in $C_{\rm end}$, but rather the original curve configuration $C$.
To this end, we reiterate that $B_{2,1}$ obtained by the contraction of the $(-1)$-curves in $C$ is smooth. A smooth point of a complex surface is an \emph{at worst terminal singularity} and therefore again rational. Therefore we can apply  (\ref{rdfa})  to the contraction map (\ref{rhomap}) and conclude
\bea
H^i(C, {\cal O}) = 0, \qquad i> 0 \,.
\eea

Suppose now that $\bar K$ is contractible on the base $B_2$. We have just shown that this implies 
$H^1 (\bar K, \cO)=0$, and what remains is to argue that this is in contradiction with $\bar K$ being the anti-canonical divisor of $B_2$.
To this end we invoke the Riemann-Roch theorem for an arbitrary curve $C$ embedded in the surface $B_2$, 
\beq\label{rr}
(K+C)\cdot C = 2 p_a(C)-2 \ . 
\eeq
The arithmetic genus $p_a(C)$ is given as
\beq\label{ag}
p_a(C)= 1-h^0(C, \cO_C) + h^1(C, \cO_C) \ . 
\eeq 
For a smooth and irreducible curve $C$, the arithmetic genus $p_a(C)$ equals the geometric genus $g(C)$, but more generally it is $p_a(C)$ which is well-defined for an arbitrary curve $C$; likewise (\ref{rr}) holds even for singular or reducible curves.

Upon applying eq.~\eqref{rr} to the anti-canonical divisor $C = \bar K$, we immediately see that
\beq \label{pa1}
p_a(\bar K) = 1 \ . 
\eeq
If $\bar K$ is smooth and irreducible, this reduces to the well-known statement that the anti-canonical divisor has (geometric) genus one, as expected because by adjunction $c_1({\bar K}) = 0$. But in general $\bar K$ is reducible and in particular non-smooth, and in such a situation the correct statement is (\ref{pa1}).
In any event, if $\bar K$ were contractible and hence $H^1 (\bar K, \cO)=0$, then the arithmetic genus of $\bar K$ would have to obey
\beq
p_a(\bar K) = 1 - h^0(\bar K, \cO) + h^1(\bar K, \cO) =1-h^0(\bar K, \cO) \leq 0 \ ,  
\eeq
which contradicts eq.~\eqref{pa1}. 

This concludes our geometric proof that the height pairing $\Ub_A$ is not contractible even for a general surface base $B_2$ for F-theory models, whether or not there exist non-abelian gauge group factors.

\section{6d SCFTs with abelian flavour symmetries} \label{sec_examples}

As mentioned already in section \ref{sec_geomintF}, the proper way to realize the decoupling limit in F-theory is to take the volume of the base to infinity while keeping some of the curves wrapped by non-abelian 7-brane stacks of finite size. The gauge theories obtained after this first operation
 are expected to flow to a non-trivial $N=(1,0)$ SCFT if we further shrink, in a second step, the wrapped curves to zero volume within the non-compact base. 
 In this sense, the field  theory arising after step one is typically viewed as an SCFT on its tensor branch in the literature.
We have shown that upon decoupling gravity from a compact F-theory model, 't Hooft anomalies arise for the abelian global symmetries while the ABJ anomalies are absent. This implies that the abelian symmetry survives as a flavour symmetry of the field theory arising after step one. The anomaly polynomial for the 't Hooft anomalies of these flavour symmetries is an important characteristic feature of the theory.
 It is furthermore expected that the abelian flavour symmetries which arise after step one persist as non-trivial abelian flavour symmetries of the interacting SCFT.
 Since moving into the tensor branch only breaks conformal invariance, we may compute the 't Hooft anomaly polynomial of the SCFT on the tensor branch~\cite{Ohmori:2014kda, Intriligator:2014eaa}.

A systematic construction of such SCFTs with abelian flavour symmetry along the tensor branch within F-theory involves several aspects, the first two of which have already been accomplished:
\begin{enumerate}
\item Classify the possible configurations of shrinkable curves in a local F-theory base. These are the curves which can support non-trivial gauge symmetries in the limit of decoupling gravity after step one.
This has been accomplished in \cite{Heckman:2013pva}, which has shown that all such curve configurations are the blowup of a  $\mathbb C^2/\Gamma$ singularity for $\Gamma \subset U(2)$ a discrete subgroup.
Note that the classification is local in the sense that it is not guaranteed (nor required) that each such configuration has an embedding into a \emph{compact} F-theory base. In particular, the list of local curve  configurations includes infinite chains. 

To each of these curve configurations one associates a minimal gauge group, which is the gauge group that cannot be higgsed any further \cite{Morrison:2012np}.
These minimal gauge theories contain, by definition, no charged hypermultiplets, which would open up a Higgs branch, but at best half-hypermultiplets in pseudo-real representations. 
Geometrically the half-hypermutliplets sit at fixed, separate locations of the curve configuration.
Such minimal models cannot be decorated by abelian flavour symmetries in a non-trivial way without changing the non-abelian gauge symmetries as only full hypermutliplets can occur in a complex representation and hence acquire a $U(1)_F$ charge.\footnote{In the minimal models it is also not possible to pair up half-hypermultiplets located at different points into full hypermultiplets. 
Pairing up, say, two half-hypermultiplets in a pseudo-real representation into a full hyper is equivalent to viewing the two half-hypers as transforming as a ${\bf 2}$ of a global ${\frak so}(2)_F$, which shows that they can acquire $U(1)$ charge.}

\item
Classify the possible non-abelian gauge enhancements of the minimal gauge theories. This has been achieved in \cite{Heckman:2015bfa}. The list includes for instance theories on single shrinkable curves and their non-abelian enhancements, and a number of chains of curves together with their possible non-abelian gauge enhancements and the resulting charged matter. 
Again, a necessary condition to decorate these models with abelian flavour symmetries is the existence of matter hypermultiplets. 
(This includes the appearance of more than one half-hypermutliplet in a given pseudo-real representation of the non-abelian gauge group, which can acquire charge if the position of the half-hypers can be tuned to coincide such that they pair up to full hypermultiplets.) 

\end{enumerate}

Before coming now to the possible abelian flavour symmetries in this list of models, 
let us first recall the status of non-abelian global symmetries. We must distinguish between two different notions of such symmetries.  
\begin{itemize}
\item
The field theoretic non-abelian global symmetry on the tensor branch can be read off directly from the spectrum:
This part of the global symmetry is $\mathfrak{su}(N)$ acting on $N$ hypermultiplets in a complex representation of the gauge group, $\mathfrak{so}(2N)$ acting on $N$ such hypers in a quaternionic representation and $\mathfrak{sp}(N)$ acting on $N$ hypermultiplets in a real representation (see e.g. \cite{Bertolini:2015bwa}). 
Strictly speaking one should distinguish between the global symmetry along the tensor branch and in the SCFT. 
{According to the general lore, moving onto the tensor branch of an SCFT only breaks conformal symmetry.}
To our understanding, it is only in one case that the above `naive' non-abelian global symmetry along the tensor branch is known to differ from the latter, where an $\mathfrak{so}(7)$ at the SCFT point is enhanced to $\mathfrak{so}(8)$ along the tensor branch \cite{Ohmori:2015pia} (see also the discussion in \cite{Bertolini:2015bwa,Heckman:2016ssk}). We will henceforth not distinguish between the two, even though strictly speaking we cannot rule out surprises at the origin of the tensor branch.

\item
The field theoretic non-abelian  global symmetry is to be distinguished from 
the non-abelian part of the geometrically realised global symmetry in an F-theory realisation. The latter is determined by constructing the maximal gauge group on a non-compact component of the discriminant intersecting the compact gauge curve at the location of the hypermultiplets of a given representation \cite{Bertolini:2015bwa}.\footnote{In a generic F-theory model realising the gauge symmetry, it might be that this global symmetry is (partially) broken because the residual part of the discriminant intersects the curve in distinct points. The question is then which maximal enhancement of the global symmetry group is possible by tuning the Weierstrass model without changing the gauge group. The maximally tunable model then determines the geometric non-abelian global symmetry.} 

\end{itemize}
In all examples studied in the literature, the {maximal}  non-abelian geometric global symmetry is contained in the field theoretic one \cite{Bertolini:2015bwa,Heckman:2016ssk}.
This is also plausible from a geometric perspective: The distinction between various configurations in F-theory with different flavour groups is due to different intersection patterns of the discriminant with the 7-brane carrying the gauge theory. In the limit of shrinking curve volume these intersection points coincide and the difference between the individual configurations is washed out.

Note that, starting from a theory with a certain  field theoretic global symmetry, a new theory can be constructed by allowing for T-brane data characterized by nilpotent orbits within the global symmetry group \cite{Heckman:2016ssk}.
This corresponds to a Higgsing process which changes in general both the global and the gauge symmetry. The non-abelian global symmetries obtained as a result of this operation matched, in all examples studied in \cite{Heckman:2016ssk}, the non-abelian part of the global field theoretic symmetry. In addition, it gives rise to abelian global symmetries, but the charges of the fields under this symmetry are not determined in \cite{Heckman:2016ssk} and the geometric origin is not worked out.

Compared to the global field theoretic symmetries as determined via the rule above ($\mathfrak{su}(N)$ versus $\mathfrak{so}(2N)$ versus $\mathfrak{sp}(N)$) abelian global symmetries are a new feature.
As the simplest example consider the case of a single hypermultiplet in a complex representation. The `non-abelian' global symmetry is $\mathfrak{su}(1)$, which is trivial at the continuous level,
 but clearly this does not preclude the possibility of the hypermutliplet carrying in addition charge with respect to a global $U(1)_F$ symmetry. We will in fact construct explicit examples of this kind in \autoref{sec_Globalmodel1} and \autoref{sec_Globalmodel2}. Similar types of charge assignments are possible enhancing for instance $\mathfrak{su}(N)$ to $\mathfrak{su}(N) \oplus \mathfrak{u}(1)_F$ for $N >1$.
 A priori, the abelian flavour factors we obtain in these F-theory examples classify as geometrically realized symmetries in the above sense. The question is then what are the maximally possible geometric symmetries that can be engineered in this way, and which field theory global symmetry do they correspond to.  More precisely, we are focusing in this work on those flavour symmetries which can be detected via their action on the 6d hypermultiplet sector.
 We will make a proposal for their form in \autoref{subsec_fieldexp}.

Recently, \cite{Anderson:2018heq} has constructed examples of $N=(2,0)$ sectors on shrinkable curves on compact bases which carry charges under discrete and abelian gauge symmetries. 
These theories hence flow to $N=(2,0)$ conformal matter sectors with abelian flavour charges.\footnote{We notice that, despite their name, these theories actually have $(1,0)$ supersymmetry. The name stems from the fact that the shrinkable curves have self-intersection $-2$, and in absence of further tunings this theory would flow to a $(2,0)$ SCFT.}
In this section, we will present two examples of a top-down construction and one example of a possible bottom-up approach to study  $N=(1,0)$ SCFTs on their tensor branch with $U(1)_F$ flavour.
In the first approach we start with a compact base containing one (\autoref{sec_Globalmodel1}) or several (\autoref{sec_Globalmodel2}) shrinkable curves and engineer a non-abelian gauge group over them together with a gauge symmetry $U(1)_A$. We then take the decoupling limit and compute the 't Hooft anomalies first from a purely field theoretic perspective. This generalizes the method of \cite{Ohmori:2014kda} by the inclusion of abelian flavour symmetries. The resulting anomaly polynomial agrees with the anomaly polynomial of the compact model upon discarding the contribution from the decoupled fields. This illustrates our general results of section \ref{Anomaliesdecoupling} and exemplifies explicitly the absence of ABJ anomalies for the flavour symmetry.
In \autoref{subsec_fieldexp} we identify the field theoretic origin of the geometrically realized $U(1)$s of  \autoref{sec_Globalmodel1} as an anomaly-free linear combination  of massive diagonal and Cartan flavour $U(1)$s and state the expected field theoretic form of the flavour symmetry. We also present a proposal to generalize this reasoning and support it with another example.
In the bottom-up approach of \autoref{sec_localmodel} we start with one of the local curve configurations of \cite{Heckman:2015bfa} with enhanced gauge symmetry and then constrain the possible global completions of this model under certain assumptions.

\subsection{Global model with $T=1$}
\label{sec_Globalmodel1}
As our first example, let us consider the del Pezzo surface $B_2=dP_1$ as the base manifold, which is obtained by blowing up $\IP^2$ at a point. The cohomology $H^{1,1}(B_2, \IZ) \simeq \IZ^2$ is spanned by the curve classes $\left[C_\alpha\right]$ for $C_{\alpha=0,1}=L, E$, where $L$ is a hyperplane of  $\IP^2$ and $E$ is the exceptional divisor. The curves $C_\alpha$ have the intersection matrix
\bea\label{im1}
\Omega_{\alpha\beta}\equiv C_\alpha \cdot C_\beta = \left( \begin{matrix}  1& 0 \\ 0 & -1  \end{matrix} \right) \,
\eea
and the anti-canonical class of $B_2$ is given by
\beq
\left[\bar K\right]=3\left[L\right]-\left[E\right] \,.
\eeq

The curve $C_1$ is hence an example of an isolated $(-1)$-curve. The minimal gauge configuration over such a curve is the trivial one \cite{Morrison:2012np}, but the restriction of the elliptic fibration to $C_1$
is non-trivial and necessarily degenerates at the twelve intersection points of the discriminant $\Delta$ with $C_1$ (because $[\Delta] = 12 \left[\bar K\right]$ and $\bar K \cdot C_1 = 1$). 
If the volume of $C_1$ is taken to zero after decoupling gravity, the theory flows to the $N=(1,0)$ E-string SCFT. This theory possesses no charged 6d hypermultiplets because the $(-1)$ curve is not wrapped by a 7-brane, but it nonetheless  exhibits an $E_8$ flavour symmetry in the SCFT limit acting on the modes of the tensionless string \cite{Ganor:1996mu,Morrison:1996pp,Witten:1996qb,Ganor:1996gu,Klemm:1996hh}. In the sequel, we will be focussing only on the flavour symmetry to the extent that it is detectable in the 6d hypermultiplet sector.\footnote{In particular, the sector acting on the tensionless string modes may contain abelian factors e.g. if the particular configuration breaks the $E_8$ flavour to a subgroup with a $U(1)$ commutant \cite{Heckman:2016ssk} (see also \cite{Klemm:1996hh}). We do not consider these in our work.} The appearance of such flavour symmetry requires an enhancement of the gauge theory along the $(-1)$ curve. 

The possible gauge enhancements of this theory follow already from anomaly cancellation and are in fact listed in \cite{Heckman:2016ssk}.
As a simple example we begin with an $SU(N)$ enhancement for $N=5$ and augment it by a $U(1)$ gauge symmetry over the compact base $B_2$.\footnote{Our motivation to choose the gauge algebra  $\mathfrak{su}(5)$ is because this is the simplest example over a $(-1)$ curve in which the theory contains a complex representation with only a single hypermuliplet, see (\ref{m3}). This will be of some heuristic value in the next section. \label{su5afootnote}} 
In this section, we will  study the resulting $U(1)$ first from the perspective of the compact geometry and then discuss the decoupling limit. This illustrates the abstract discussion of \autoref{sec_U1gaugsymmetries}. An interpretation of the $U(1)$
from the more general field theory perspective outlined at the beginning of this section will be given in \autoref{subsec_fieldexp}.

The simplest realisation of a Weierstrass fibration with a single $U(1)$ is given by the $U(1)$ restricted Tate model \cite{ Grimm:2010ez}, which corresponds to a fibration in Tate form
\bea
y^2 + a_1 x y z  + a_3  y  z^3 = a_2 x^2  z^2 + a_4  x z^4 + a_6 z^6  \, ,
\eea 
with  $a_6 = 0$. Here the fiber coordinates $\left[x : y : z\right]$ are homogenous coordinates of $\mathbb P_{231}^2$ and $a_m$ are global sections of the line bundle $\cO_{B_2}(m\bar K)$. The extra section $s_A$ sits at $\left[0 : 0 : 1\right]$ and intersects the singularity in the $I_2$ fiber over $\{ a_3 = 0\} \cap \{ a_4=0 \}$. Resolving this singularity leads to a toric blow-up divisor $S_A$ which is identified with the section divisor ${\rm div}(s_A)$.
The inclusion of non-abelian gauge algebras is possible by restricting the $a_m$ following Tate's algorithm \cite{Bershadsky:1996nh}. 
Ignoring the $U(1)$ for a second, we engineer a non-abelian gauge algebra $SU(5)$
 to be supported on the $(-1)$-curve 
\beq
C_1: \mathfrak{su}(5) 
\eeq
by setting
$a_2 = a_{2,1} w$, $a_3 = a_{3,2} w^2$, $a_4 = a_{4,3} w^3$, $a_6 = a_{6,5} w^5$
with $C_1 = \{w=0 \}$, followed by a resolution of the $I_5$ singularity in fiber over $w=0$ \cite{Krause:2011xj}.
The discriminant of the Weierstrass model takes the form
\bea \label{DeltaSU5U1}
\Delta = \frac{1}{16} w^5 (a_1^4 \, P + {\cal O}(w)) \, \qquad P = a_{2,1} a_{3,2}^2 - a_1 a_{3,2} a_{4,3} + a_1^2 a_{6,5}   \,.
\eea
Setting $a_{6,5} = 0$ engineers an extra $U(1)$ gauge group factor and leads to a factorization of the polynomial $P$ as $a_{3,2} (a_{2,1} a_{3,2} - a_1 a_{4,3})$.

The non-abelian anomaly coefficient is
\beq
b_{\mathfrak{su}(5)} = (0,1)\,,
\eeq
where $\left[C_\alpha\right]_{\alpha=0,1}$ have been used as the basis elements of $H^{1,1}(B_2, \IZ)$. Since in the current model $S_A \cdot Z = 0$ and $\pi_\ast(S_A \cdot E_i) = \delta_{i3} \, C_1$ with $E_i$ the $\mathfrak{su}(5)$ resolution divisors, 
the height pairing for $U(1)_A$ is readily computed as
\beq
b_A= 2 \bar K - \frac65 C_1 = (6,-\frac{16}{5})\,.
\eeq
There are three types of $\frak{su}(5)$ charged matter fields, localized at the intersection of $C_1$ with the curve $\{a_{3,2}=0\}$, $\{ a_{2,1} a_{3,2} - a_1 a_{4,3} =0\}$ and $\{a_1 = 0\}$ in the base. Their  respective charges ~\cite{Krause:2011xj} and multiplicities are
\bea\label{m1}
\bold{5}_{3/5}&:&  C_1 \cdot (3\bar K - 2 C_1) = 5\,,\\ \label{m2}
\bold{5}_{-2/5}&:& C_1 \cdot (5\bar K - 3 C_1)  = 8\,,\\ \label{m3}
\bold{10}_{1/5}&:& C_1 \cdot \bar K  = 1 \,, \\ \label{m4}
\bold{1}_{1}&:& (4\bar K - 3C_1) \cdot (3 \bar K - 2 C_1) = 73 \,.
\eea
Finally, there arise $121$ neutral hypermultiplets  away from $C_1$ at points where $a_{4,3=0}$ and $a_{3,2}=0$ that are not charged under the non-abelian group $G$ nor the $U(1)_A$. 

One can check that this spectrum is anomaly-free before taking the decoupling limit. Upon decoupling gravity, we may compute the anomaly polynomial of the resulting theory by discarding the contribution of the gravity multiplet together with the abelian vector and the hypermultiplets that are uncharged under the non-abelian gauge group. The one-loop anomaly due to the remaining multiplets is described by the anomaly polynomial
\bea\nn
I^{\rm one-loop}|_{\rm local}&=&I^{\rm tensor} + I^{\rm vector} + I^{\rm hyper}
\eea
with
\bea
I^{\rm tensor}&=& \frac{29}{5760} ({\rm tr} R^4 + \frac54({\rm tr}R^2)^2)  -\frac{1}{128}({\rm tr}R^2)^2 \\
I^{\rm vector} &=& I^{\rm vector}_{{SU(5)}}  \\ 
&=&- \frac{24}{5760} ({\rm tr} R^4 + \frac54({\rm tr}R^2)^2) -\frac{1}{24}( 10\,{\rm tr} F_{G}^4 +6({\rm tr} F_{G}^2)^2 )+ \frac{10}{96} {\rm tr} F_{G}^2 {\rm tr}R^2 \\ \nn
I^{\rm hyper} &=& 5 I^{\rm hyper}_{\bold 5_{3/5}} + 8I^{\rm hyper}_{\bold 5_{-2/5}} +I^{\rm hyper}_{\bold{10}_{1/5}} \\
&=& \frac{75}{5760}  ({\rm tr}R^4 + \frac{5}{4} ({\rm tr} R^2)^2) + \frac{1}{24}(10 {\rm tr} F^4_G + 3({\rm tr} F_G^2)^2 +\frac{480}{25}{\rm tr}F_G^2F_{A}^2 +\frac{107}{25}F_{A}^4 )  \\
&& -\frac{1}{96}(16 {\rm tr}F_{G}^2 + \frac{395}{25} F_A^2 ){\rm tr} R^2    \,.
\eea
$F_G$ and $F_A$ denote the non-abelian and abelian field strengths, respectively, and the expressions follow from the general anomaly polynomials collected in \autoref{sec_anomalies}. We stress that the above expression is valid in the decoupling limit, where $F_A$ and $R$ are only background fields, while $F_G$ is dynamical.

The GS contribution is uniquely determined by requiring the theory to be free of all anomalies involving the non-abelian gauge field, which yields
\beq\label{model1GS}
I^{\rm GS}|_{\rm local} = \frac{1}{8} ({\rm tr} F_G^2 + \frac{1}{4} {\rm tr}R^2 - \frac{16}{5}F_A^2)^2 \,.
\eeq
This is analogous to the procedure in \cite{Ohmori:2014kda}, which however did not consider the possibility of abelian flavour symmetries.

As we now show, the same GS contribution can be obtained purely from the F-theory geometry following the general discussion of \autoref{sec_U1gaugsymmetries}. Indeed, upon decoupling gravity we may compute the GS term in the anomaly polynomial by using \eqref{I8GSSW}  and \eqref{X4alpha} 
once we discard the contribution due to the tensor in the gravity multiplet since it decouples. This limit is obtained by taking $j^1\to 0$ and $j^0\to 1$ which shrinks the $(-1)$-curve keeping the total volume fixed. In this particular limit, the duality matrix \eqref{Dmatrix} reduces to
\begin{align}
D(j) = \left (\begin{array}{cc} 2(j^0)^2-1 & -2j^0 j^1 \\ 2j^0 j^1 & -2(j^1)^2-1 \end{array}\right ) \to \left (\begin{array}{cc} 1 & 0 \\ 0 & -1 \end{array}\right ) \,.
\end{align}
This means that the tensor in the gravity multiplet, which is self-dual, is along the direction of $j=(1,0)$ in the decoupling limit. Similarly, the anti-self-dual tensor which remains in the SCFT is along $b_{\mathfrak{su(5)}} = (0,1)$. Thus, in order to remove the contribution due to the gravity multiplet, we simply replace the anomaly coefficients $a =K$ and $b_A$ entering the GS counterterms via \eqref{I8GSSW}  and \eqref{X4alpha}  as in
\bea
a = (-3,1) &\to& a_\parallel = (0,1) \\
b_A = (6, -\frac{16}{5}) &\to& b_{A \parallel} = (0,-\frac{16}{5}) \ . 
\eea
Using these projections along $b_{\mathfrak{su}(5)}$ as the new values for $a$ and $b_A$ in \eqref{I8GSSW}  and \eqref{X4alpha} we do indeed reproduce \eqref{model1GS}, hence deriving the absence of gauge anomalies even after taking the decoupling limit. 

The total anomaly polynomial after decoupling gravity is therefore
\bea\nn
I^{\rm tot}|_{\rm local}&=&(I^{\rm one-loop} + I^{\rm GS})|_{\rm local} \\
&=&  \frac{1}{72} ({\rm tr} R^4 + \frac54({\rm tr}R^2)^2)+\frac{35}{24} F^4_A -\frac{35}{96} F^2_A {\rm tr}R^2 \,.
\eea
It describes the 't Hooft anomalies of the $SU(5)$ gauge theory with $U(1)_F = U(1)_A$ flavour symmetry after the decoupling.

\subsection{Field theoretic interpretation and general pattern} \label{subsec_fieldexp}

We will now understand the abelian flavour symmetry found in the previous model in more detail from a field theoretic perspective.
Consider first a generic Tate model realising the $G = \mathfrak{su}(5)$ gauge algebra over $C_1$ by setting $a_6 = a_{6,5} w^5$, $a_{6,5} \neq 0$.
There are $13$ hypermultiplets in the ${\bf 5}$ representation of $G$, located at the $13$ zeroes of the polynomial $P$ in ({\ref{DeltaSU5U1}) along $w=0$.
For generic values of the moduli, these $13$ points are distinct and the {\it manifest} geometrically realized non-abelian flavour symmetry is trivial. However, at special values of the complex structure moduli the non-abelian flavour group as realized in the F-theory model can be enhanced up to the maximal geometrically realizable non-abelian group $SU(13)$, in which case in particular all of the $N=13$ ${\bf 5}$-hypermultiplets localize at the same point on $C_1$.
This matches with field theoretic expectation for the non-abelian part of the flavour group, which 
is $SU(13) \times SU(1) \equiv SU(13)$, acting on the $N=13$ hypermultiplets in the ${\bf 5}$ representation of $G$, while the ${\bf 10}$ is an $SU(13)$ singlet.
In particular this is in agreement with the observation of \cite{Bertolini:2015bwa,Heckman:2016ssk} that in any given F-theory realisation, the maximal geometrically realised flavour group is contained in the field theoretic one.

The fact that the ${\bf 10}$ hypermultiplet is an $SU(13)$ singlet makes it particularly obvious that 
the abelian $U(1)_A$ factor found by setting $a_6=0$ cannot be embedded into this non-abelian $SU(13)$ flavour group (see footnote \ref{su5afootnote}).
We conclude that the rule of assigning to $N$ complex hypermultiplets a flavour symmetry factor $SU(N)$ misses this possibility of extra abelian flavour symmetries.

The mismatch can be remedied by noting that a priori in field theory
$N$ hypermultiplets should be acted on by a flavour group $U(N)$ rather than $SU(N)$. This would give a field theoretic flavour group, in the generic model with $a_6 \neq 0$, of the form\footnote{We are ignoring here the global structure of the group, writing for simplicity  $U(13) = SU(13) \times U(1)_a$.}
\bea \label{U13}
G^{(\rm trial)}_{F, {\rm gen}} = [SU(13) \times U(1)_a] \times U(1)_b  \,,
\eea
and we would assemble the spectrum in representations
\bea \label{chargesU1aU1b}
({\bf 13}_{(1,0)}, {\bf 5}), \qquad ({\bf 1}_{(0,1)}, {\bf 10}) \,.
\eea
However, the two diagonal $U(1)_a$ and $U(1)_b$ factors by themselves have a 1-loop mixed $U(1)_i  - SU(5)^3$ anomaly, which is proportional to 
\bea
{\cal A}^{\rm 1-loop}_{U(1)_A - G_\kappa^3}  = \sum_I \mathcal M_I^\kappa \, E^I_\kappa \, q_{IA} \,, \label{cubicmixed-m}
\eea
where $M_I^\kappa$ gives the multiplicities of the charged hypermultiplets in representation ${\bf R}_I$ of the non-abelian gauge group $G_{\kappa}$ with $U(1)_A$ charge $q_{IA}$.
In a compact brane construction in Type IIB language, this anomaly is cancelled by a version of the Green-Schwarz mechanism different from the one considered in the rest of this article:
It is associated with the non-trivial gauging of the axionic scalars obtained from the RR 2-form $C_2$ rather than from $C_4$. As a result,
the gauge bosons associated with the two diagonal factors $U(1)_a$ and $U(1)_b$ both acquire a mass via a geometric St\"uckelberg mechanism (as do the involved axions), as detailed in the context of four-dimensional F-theory compactifications in \cite{Grimm:2011tb}. In F-theory language the diagonal $U(1)$ gauge bosons are realized in terms of non-harmonic two-forms \cite{Grimm:2011tb,Martucci:2015dxa }, as opposed to the harmonic two-forms due to rational sections of the fibration. This makes their quantitative analysis rather involved.
In Type IIB compactifications such massive $U(1)$s appear as perturbative global symmetries, which are broken non-perturbatively: The relevant effects are D-brane instantons \cite{Blumenhagen:2006xt,Ibanez:2006da,Florea:2006si} carrying D1-charge \cite{Grimm:2011dj,Martucci:2015oaa}. They break the massive $U(1)$ either completely or to a discrete subgroup $\mathbb Z_k$ with $k >1$ \cite{Camara:2011jg,BerasaluceGonzalez:2011wy}.\footnote{The latter is automatically free of anomalies. The non-trivial $\mathbb Z_k$, $k >1$, remnant survives theoretically as a global discrete symmetry and geometrically, in F-theory, in the form of torsion homology in the elliptic fibration \cite{Camara:2011jg,Mayrhofer:2014laa}. F-theory compactifications with discrete symmetries have been studied intensively, beginning with \cite{Braun:2014oya,Morrison:2014era, Anderson:2014yva,Klevers:2014bqa,Garcia-Etxebarria:2014qua,Mayrhofer:2014haa,Mayrhofer:2014laa,Cvetic:2015moa,Kimura:2016crs}.} The first case is in fact a special instance of a $\mathbb Z_k$ symmetry with $k=1$, but it does not correspond to a global symmetry once brane instanton effects are taken into account. 

While the non-perturbative breaking of the massive $U(1)$ has been primarily studied in string compactifications to four dimensions, we expect on general grounds that the same logic applies in 6d.
Hence, we are operating under the assumption that, as far as ${\it continuous}$ global symmetries are concerned, we can ignore all combinations of potential abelian flavour symmetries for which (\ref{cubicmixed-m}) is non-vanishing. On the other hand, they are expected to play a role for ${\it discrete}$ global symmetries, possibly even of the type considered recently  in the context of 6d SCFT in \cite{Anderson:2018heq}.

To come back to our example, even though $U(1)_a$ and $U(1)_b$ are massive by themselves, suitable linear combinations can remain as massless gauge symmetries in a global F-theory model \cite{Krause:2012yh,MayorgaPena:2017eda}. 
Upon decoupling gravity, these become exact global symmetries.  
A possible relation between abelian flavour symmetries in 6d SCFTs and massive $U(1)$s has also been mentioned generally in \cite{Heckman:2016ssk}.
As discussed, the condition for masslessness is that the mixed cubic anomaly (\ref{cubicmixed-m}) vanishes at the 1-loop level so that there is no need for a mass inducing  GS term.
For the charge assignments (\ref{chargesU1aU1b}) the generator associated with this linear combination is 
\bea
T_m =  - \frac{1}{13} T_a +   T_b \,.
\eea

In a given F-theory realisation this linear combination of $U(1)_a$ and $U(1)_b$ can receive an admixture from a Cartan $U(1)_c$ within the maximal possible non-abelian flavour group - here $SU(13)$.
In the present example, such an interpretation is indeed possible.
In fact, the naive non-abelian flavour symmetry that is suggested by the charges (\ref{m1})--(\ref{m3}) is not the full $SU(13)$, but only an $SU(8) \times SU(5)$ subgroup.
If we consider the branching
\bea
SU(13) &\rightarrow&  SU(5) \times SU(8) \times U(1)_c \\
{\bf 13} &\rightarrow&  ({\bf 5},{\bf 1})_{8} \oplus ({\bf 1},{\bf 8})_{-5} \\
{\bf 1} &\rightarrow&    ({\bf 1}, {\bf 1})_0  
\eea
then the charge assignments (\ref{m1})--(\ref{m3}) can be understood by identifying the generator of the flavour group $U(1)_F = U(1)_A$ with the linear combination
\bea
T_F =  \frac{1}{5} T_m + \frac{1}{13} T_c \,.   
\eea

This admixture of the Cartan $U(1)_c$ obscures the true nature of the abelian flavour symmetry. Being a Cartan of the maximal possible non-abelian flavour symmetry, $U(1)_c$ is free not only of the cubic ABJ anomaly, but also of the quadratic ABJ anomaly, whose 1-loop piece is cancelled by the conventional GS mechanism. Furthermore, the mixed $U(1)_c - U(1)_m-SU(5)^2$ ABJ anomalies vanish at the 1-loop level, as does the associated GS term. Therefore the fact that both $U(1)_c$ and $U(1)_F$ are free of all ABJ anomalies ensures that also $U(1)_m$ is free of ABJ anomalies in the local limit.

 The realisation of the global F-theory model as a $U(1)$ restricted Tate model is only a special case of the more general Morrison-Park model \cite{Morrison:2012ei}, which in principle allows for very different charge assignments \cite{Mayrhofer:2012zy, Borchmann:2013jwa,Borchmann:2013hta,Kuntzler:2014ila}. Had we started with one of these,
 the abelian gauge symmetry in the compact model, as far as their action of the $\mathfrak{su}(5)$ charged matter is concerned, would be a linear combination of $T_m$ and a different Cartan $U(1)_c$ within $SU(13)$.
 Locally, after decoupling gravity, all these models become indistinguishable, as we will elaborate on in more detail below. 
 This shows that the maximal possible global symmetry in field theory is
\bea \label{GFSU13}
G_{F} = SU(13) \times U(1)_m \,.
\eea

Note that we do not have an example of a globally defined fibration in which the $U(1)_m$ is guaranteed to survive as a massless $U(1)$.\footnote{Interestingly, the required combination of charges $q_m = -1$ for ${\bf 5}$ and $q_m=13$ for ${\bf 10}$ does appear in the list provided by \cite{Lawrie:2015hia} for {\it in principle} compatible charges in $\mathfrak{su}(5)$ models with non-singular sections, namely the configuration $I_5^{(0|1)}$. A global realisation as a canonical Morrison-Park model seems not possible on the base $B_2$, as follows by working out the constraints implied by  the vanishing orders determined in \cite{Kuntzler:2014ila}. To the best of our understanding, this does not yet preclude the existence of non-canonical models.}
 Fortunately, this is not required to support the proposal (\ref{GFSU13}). To appreciate this, consider again the most generic $\mathfrak{su}(5)$ Tate model by taking $a_{6,5} \neq 0$. Globally this model does not support a massless $U(1)$. But as is well known, we can view this model as a $U(1)$ restricted model after Higgsing the $U(1)_A$. The Higgsing occurs as a conifold transition at the points $a_{3,2} = a_{4,3} = 0$ away from the $\mathfrak{su}(5)$ curve. This is the locus of the charged $\mathfrak{su}(5)$ singlets, which act as the Higgs fields.
In the decoupling limit the information about whether or not the Higgsing has been performed is not available any more because the singlets are decoupled from the $\mathfrak{su}(5)$ field theory. Geometrically, they are infinitely far away after scaling up the directions normal to the compact curve $C_1$. 
A finite non-zero vacuum expectation value of the Higgs fields, which breaks the $U(1)_A$ as a gauge symmetry, is washed out from the perspective of the $\mathfrak{su}(5)$ field theory in the decoupling limit.
 One can phrase the same phenomenon geometrically: The extra section  of the $U(1)$ restricted Tate model at $[x : y : z] = [0 : 0: 1]$ continues to exist {\it locally} in the vicinity of the curve $C_1$ at $w=0$ even after deforming the model by letting $a_{6,5} \neq 0$. Globally, this destroys the rational section due to the appearance of branch cuts originating at the locus $a_{3,2} = a_{4,3} = 0$, but locally and away from this branching locus the model is unaffected by the deformation.
In this sense, even the most generic $\mathfrak{su}(5)$ Tate model gives rise to a $U(1)$ flavour symmetry after decoupling gravity - albeit one that cannot be extended globally. 
Combined with the fact that in the generic Tate model, the global non-abelian symmetry $SU(13)$ can be attained, this supports our claim.

Let us present another  example illustrating this point, in which in fact no admixture of a flavour Cartan symmetry to the flavour symmetry  occurs even at the level of the globally extended model.
To this end, we realize the gauge algebra ${\mathfrak{e}}_6$ along the $(-1)$ curve $C_1$. Anomaly cancellation implies $N=5$ hypermultiplets in the ${\bf 27}$. The maximal non-abelian flavour symmetry, including a potential diagonal $U(1)$, is hence
\bea
G_{F} = U(5) = SU(5) \times U(1)_a \,.
\eea 
Since ${\rm tr}_{\bf 27} F^3 = 0$, the diagonal $U(1)_a$ is free of the mixed cubic anomaly (\ref{cubicmixed-m}) and has therefore a chance to survive even by itself as a massless gauge symmetry in a compact model.
This is confirmed by an explicit geometric analysis: To stay in the example of the $U(1)$ restricted Tate model over $B_2$, if we engineer  the vanishing orders $a_1  = a_{1,1} w$, $a_2  = a_{2,2} w^2$, $a_3  = a_{3,2} w^2$, $a_4  = a_{4,3} w^3$ (with $C_1 = \{w = 0\}$), the discriminant of the elliptic fibration takes the form
\bea
\Delta = \frac{1}{16} w^8 ( 27 a_{3,2}^4 + {\cal O}(w)) \,. 
\eea
The $5$ hypermultiplets in the ${\bf 27}$ are localised at the intersection of $\{a_{3,2} = 0\}$ with $C_1$. An explicit resolution and analysis of the fibers shows that each of these carries  charge $q_A =-\frac{1}{3}$ with respect to a $U(1)_A$ with height pairing $b_A = 2 \bar K - \frac{4}{3} C_1$ \cite{unpublished}. The $U(1)_A$
is free of ABJ anomalies in the decoupling limit, hence giving rise to a global symmetry $SU(5) \times U(1)_A$. 
If we now break the $U(1)_A$ gauge group (while keeping the ${\mathfrak{e}}_6$)  in the globally defined model by setting $a_6 = a_{6,5 } w^5$ for $a_{6,5 } \neq 0$, the form of discriminant does not change to leading order in $w$. Again, locally, away from the singlet locus, the two models are indistinguishable. The fact that the $U(1)_A$ survives as a flavour symmetry after decoupling, on the other hand, only requires local information, in particular the vanishing of the mixed cubic anomaly (\ref{cubicmixed-m}).

To conclude this discussion, we make the following proposal to determine the field theoretic global symmetry in the decoupling limit, as far as its action on the 6d hypermultiplet sector is concerned:
 Assign to $N_i$ hypermultiplets in a complex representation ${\bf R}_i$ of the gauge group an initial flavour group factor $U(N_i) = SU(N_i) \times U(1)_i$.
 The possible abelian flavour symmetries not contained in the Cartan of the non-abelian part 
 of the flavour group are then those linear combinations $\sum_i x_i \, U(1)_i$ which are free of the mixed cubic 1-loop ABJ anomaly (\ref{cubicmixed-m}) with the non-abelian gauge group.\footnote{In the case of pseudo-real or real representations, the non-abelian part of the flavour group is enhanced from $SU(N)$ to $SO(2N)$ or $Sp(N)$, respectively. When a suitable subgroup is gauged, it might be necessary to consider a branching which leads to hypermultiplets charged under a `diagonal' $U(1)$ at intermediate stages, and these can enter the anomaly free linear combination $\sum_i x_i \, U(1)_i$. An example will be discussed in the following subsection. \label{footnote-gauging}}
In a concrete geometric realisation of the model, it can, and generically does happen that one finds a $U(1)$ given by a linear combination of these $U(1)$s and some Cartan $U(1)$ within the maximal non-abelian flavour group; naively the latter is broken, in the concrete geometric realisation, to a corresponding subgroup. However, the field theoretic global symmetry group, in particular at the SCFT point, is conjectured to be the one with the maximal non-abelian gauge group plus the additional abelian combinations $\sum_i x_i \, U(1)_i$.\footnote{Recall, however, from the discussion at the beginning of \autoref{sec_examples} that in principle the flavour symmetry at the origin of the tensor branch might be smaller than the `naive' symmetry determined away from the SCFT point.} 
 
In particular, the number of independent abelian flavour symmetries acting on the non-abelian theory in the decoupling limit is determined in terms of local data. 
This number can be bigger or smaller than the number of abelian gauge symmetries in a given global embedding of the model. In the latter case the distinction between the extra globally realized gauge symmetries is due to  the $U(1)$ charges of states uncharged under the local gauge group, for instance
extra $U(1)$  charged singlet states, which have been decoupled from the local non-abelian gauge theory.

In this sense, the construction of a global model with a $U(1)$ is only one possible global completion of the SCFT. Irrespective of whether or not a flavour $U(1)$ survives as a gauge theory in this particular global completion, the abelian part of the flavour group is as stated above.

\subsection{Global model with $T=2$}
\label{sec_Globalmodel2}
The next example involves a chain of two $(-2)$-curves. The minimal gauge theory on such curves is trivial, and the elliptic fibration over the $(-2)$-curves is just a product. 
The theory of $k$ linearly intersecting $(-2)$-curves is in fact simply the $A_k$ $N=(2,0)$ SCFT. We will again have to first enhance to a non-abelian gauge algebra, which we can then decorate by a $U(1)$ gauge symmetry and study its behaviour upon decoupling gravity, where it becomes a flavour symmetry. SCFTs over a chain of $(-2)$-curves with abelian and discrete charges have also been considered recently in \cite{Anderson:2018heq}.

 Let us start by the weighted projective space $\IP^2_{132}$, whose homogeneous coordinates $z_{k=0,1,2}$ are subject to the $\mathbb C^\ast$ identification
\beq
(z_0, z_1, z_2) \sim (\lambda z_0, \lambda^3 z_1, \lambda^2 z_2) \,, \qquad \lambda \in \mathbb C^\ast \,.
\eeq
Note that there exist two point-like orbifold singularities at $z_0=z_1=0$ and at $z_0=z_2=0$, which can be resolved by inserting the rational curves
\beq
F: \{f=0\}\,,~\qquad E_{a}: \{e_a=0\}~~\text{for}~a=1,2\,,
\eeq
respectively. Here, $f$ and $e_a$ are the additional homogeneous coordinates of the blown-up surface. We take this resolved manifold as the base of our F-theory model, 
\beq
B_2=\widehat{\IP^2_{132}} \,.
\eeq
It is a toric variety described by the fan in Figure~\ref{f:fan}.

\begin{figure}[t]
\centering
\includegraphics[scale=.6]{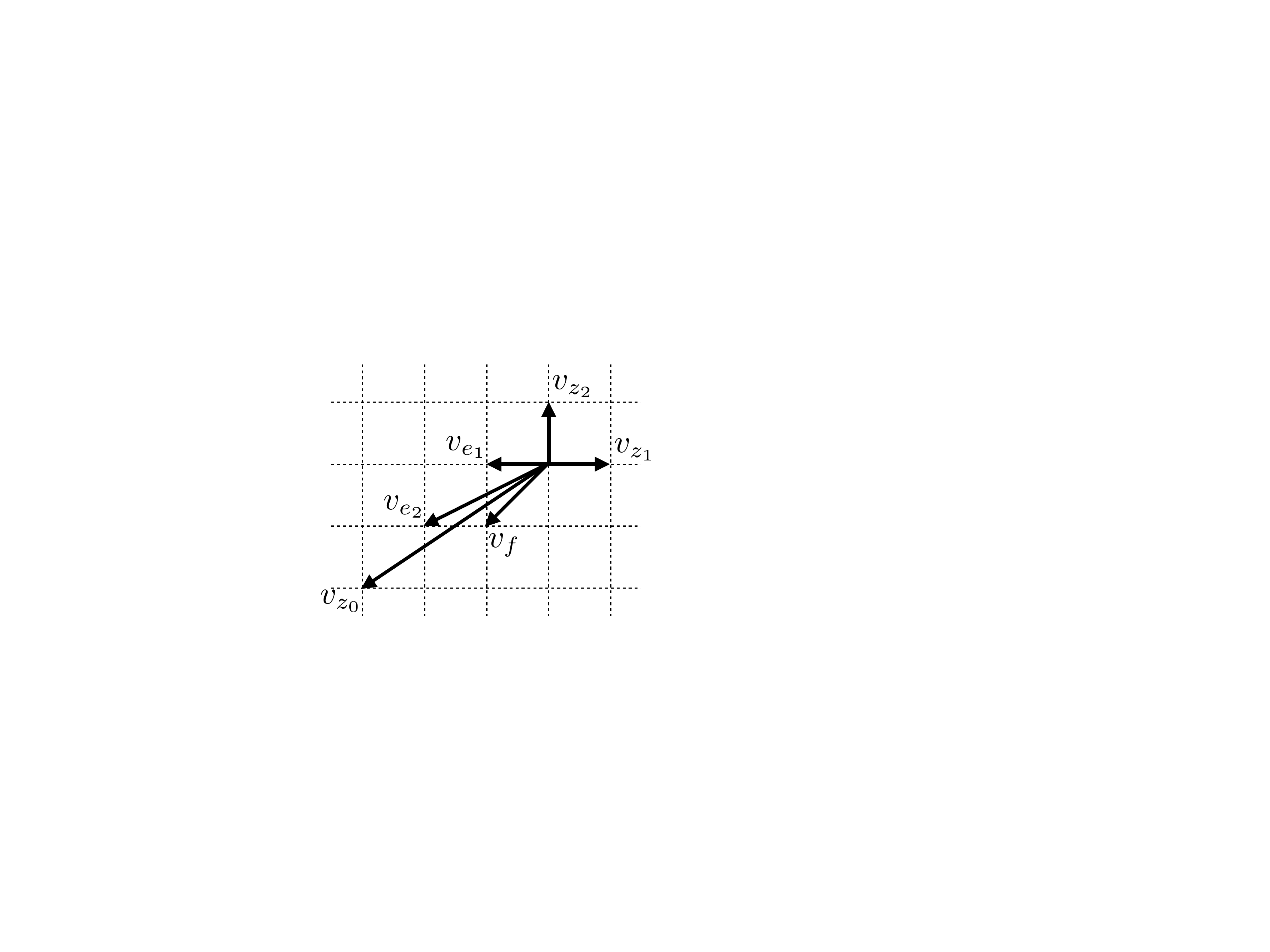}
\caption{The two-dimensional fan for the toric surface $\widehat{\IP^2_{132}}$. To each ray is associated a homogeneous coordinate indicated by the subscript.}\label{f:fan}
\end{figure}

Denoting the toric divisors $\{z_k=0\}$ by $D_{k}$ ($k=0,1,2$), one can see that the cohomology $H^{1,1}(B_2, \IZ)$ is spanned by the four curve classes $\left[C_\alpha\right]$ for $C_{\alpha=0,\dots,3}= D_0, F, E_1, E_2$ while the two toric divisor classes $\left[D_1\right]$ and $\left[D_2\right]$ are subject to the  linear equivalence relations
\bea
\left[D_1\right] &=& 3\left[D_0\right]+\left[F\right]+\left[E_1\right]+2\left[E_2\right] \,,\\
\left[D_2\right] &=& 2\left[D_0\right]+\left[F\right] +\left[E_2\right]\,.
\eea
The intersection matrix is also computed easily from the fan as
\bea\label{im}
\Omega_{\alpha\beta}\equiv C_\alpha \cdot C_\beta = \left( \begin{matrix}  -1& 1 & 0 & 1 \\ 1 & -2 & 0 & 0  \\ 0 & 0 & -2 & 1 \\ 1 & 0 & 1 & -2  \end{matrix} \right) \,,
\eea
which has one positive and three negative eigenvalues. 
Finally, the anti-canonical class is given by
\bea
\left[\bar K\right] &=& \sum_{k=0}^2 \left[D_k\right] +\left[F\right]+\sum_{a=1}^2\left[E_a\right] \\
&=& 6 \left[C_0\right] + 3\left[C_1\right] + 2\left[C_2\right] + 4\left[C_3\right] \\
&=& (6,3,2,4) \,, 
\eea
where $\left[C_\alpha \right]_{\alpha=0,\dots,3}$ have been used as the basis elements of $H^{1,1}(B_2, \IZ)$.

In absence of further enhancements, F-theory on this basis flows to the $N=(2,0)$ SCFT $A_1 \oplus A_2$, where the $A_1$ is supported on the $(-2)$-curve $F$ and the $A_2$ on the chain of $(-2)$-curves $E_1$ and $E_2$. We will engineer a non-trivial non-abelian gauge enhancement over the latter chain and augment it by an abelian gauge group.
For example, let us construct a Weierstrass model over $B_2$ for which the gauge group is $G \times U(1)_A$ with $G=SU(2) \times SU(3)$. Instead of starting from scratch, we may apply to the $B_2$ of our choice what is known about the sixteen toric hypersurface fibrations~\cite{Klevers:2014bqa}, one of which has exactly $G \times U(1)_A$ as the generic gauge group. To be more specific, we first fiber an appropriately chosen toric surface $S$ over $B_2$ and impose a hypersurface equation to obtain an elliptic curve embedded in $S$ as the generic fiber. This is done in such a way that the Mordell-Weil group is of rank one (indicating one extra independent rational section) and the fibers over the curves $E_1$ and $E_2$ degenerate to type $I_2$ and $I_3$, respectively. We compute below the relevant properties of the model by using the results in Section 3.5.1 of  \cite{Klevers:2014bqa}, to which the reader is referred for the details on the fiber geometry.

We start by engineering the non-abelian gauge algebras to be supported on the $(-2)$-curves
\beq
C_2: \mathfrak{su}(2)\,, \qquad C_3:  \mathfrak{su}(3)\,,
\eeq
so that 
\beq
b_{\mathfrak{su}(2)}=(0,0,1,0)\,,\qquad b_{\mathfrak{su}(3)}=(0,0,0,1) \,.
\eeq
In such a model, the height pairing for the $U(1)_A$ is computed, with the help of \cite{Klevers:2014bqa}, as
\beq
b_A= 2 \left[\bar K\right] - \frac12\left[C_2\right]-\frac23\left[C_3\right] = (12, 6, \frac{7}{2}, \frac{22}{3}) \,,
\eeq 
and the complete spectrum of charged matter representations is obtained from the intersection matrix~\eqref{im} of $B_2$ as follows:
\bea\label{mp1}
(\bold 2, \bold 3)_{-1/6}&:& C_3 \cdot C_2 = 1 \,, \\ \label{mp2}
(\bold 2, \bold 1)_{1/2}&:& C_2 \cdot (8\bar K -2C_2-3C_3) = 1\,, \\ \label{mp3}
(\bold 1, \bold {3})_{-2/3}&:& C_3 \cdot (3\bar K -C_2-C_3)= 1\,,\\ \label{mp4}
(\bold 1, \bold { 3})_{1/3}&:& C_3 \cdot (6\bar K -C_2-2C_3)=3 \,, \\ \label{mp5}
(\bold 1, \bold 1)_{-1}&:& (3\bar K -C_2-C_3)\cdot (4\bar K - C_2-2C_3)=69\,. 
\eea
In addition to these charged hypermultiplets there are $109$ uncharged hypers.

As expected, this spectrum satisfies all the anomaly cancellation conditions before we decouple gravity. As in the previous example, the $G$-singlets disappear upon decoupling gravity together with the gravity multiplet. Furthermore, the tensor associated to the curve $C_1$ decouples from the theory on $C_2$ and $C_3$ because the latter are disjoint from $C_1$; hence we remove its contribution when we compute the anomaly polynomial of the resulting SCFT. The one-loop anomalies of tensor, vector and hypermultiplets after decoupling are then
\bea\nn
I^{\rm one-loop}|_{\rm local}&=&I^{\rm tensor} + I^{\rm vector} + I^{\rm hyper} \\  \nn
&=&  -\frac14({\rm tr} F_{SU(2)}^2)^2 -\frac14 ({\rm tr} F_{SU(3)}^2)^2 +  \frac14 \tr F_{SU(2)}^2 \tr F_{SU(3)}^2 + \frac{1}{12} F^2 \tr F_{SU(2)}^2  \\&&\, + \frac{5}{24} F^2 \tr F_{SU(3)}^2 
+\frac{5}{144} F^4 - \frac{1}{32} F^2 \tr R^2 +\frac{67}{5760} (\tr R^4 + \frac{5}{4} (\tr R^2)^2 ) \,.
\eea
Here $I^{\rm tensor}$ is the contribution of the two remaining tensor multiplets (coupling to $C_2$ and $C_3$) and
\bea
I^{\rm vector} &=& I^{\rm vector}_{{SU(2)}} + I^{\rm vector}_{{SU(3)}} \\
I^{\rm hyper} &=& I^{\rm hyper}_{(\bold2, \bold3)_{-1/6}} +I^{\rm hyper}_{(\bold2, \bold1)_{1/2}} +I^{\rm hyper}_{(\bold1, \bold{3})_{-2/3}} +3I^{\rm hyper}_{(\bold1, \bold {3})_{1/3}} \,. 
\eea
The GS contribution after decoupling is uniquely determined from the requirement that there are no gauge anomalies \cite{Ohmori:2014kda},
\bea\nn
I^{\rm GS}|_{\rm local}&=& \frac14 (\tr F_{SU(2)}^2)^2 + \frac14 (\tr F_{SU(3)}^2)^2 -\frac14 \tr F_{SU(2)}^2 \tr F_{SU(3)}^2 -\frac{1}{12} F^2 \tr F_{SU(2)}^2 - \frac{5}{24} F^2 \tr F_{SU(3)}^2 \\ \label{gs-model2}
&&\,+\frac{13}{144} F^4  \,.
\eea
The total anomaly polynomial is hence 
\bea\nn
I^{\rm tot}|_{\rm local}&=& (I^{\rm one-loop} + I^{\rm GS})|_{\rm local} \\
&=&  \frac{18}{144}F^4 - \frac{1}{32} F^2 \tr R^2 +\frac{67}{5760} (\tr R^4 + \frac{5}{4} (\tr R^2)^2 )\,.
\eea

Let us conclude our analysis of this model by confirming that the same GS contribution arises from the F-theory geometry. In order to do this, we need to compute the new anomaly coefficient vectors $a, b_A,\,b_{\mathfrak{su}(2)}, \,b_{\mathfrak{su}(3)}$ without the contribution due to the tensors that decouple. In this particular case, these are the tensor in the gravity multiplet as well as the one associated to the curve $C_1$. In the limit where we decouple the $SU(2)\times SU(3)$ sector from the rest, i.e.~when the volume of the curves $b_{\mathfrak{su}(2)}$ and $b_{\mathfrak{su}(3)}$ goes to zero, we must  use the projection of $a$ and $b_A$ onto the subspace spanned by $\langle b_{\mathfrak{su}(2)},b_{\mathfrak{su}(3)} \rangle$. These projections are
\begin{align}
a_\parallel &= 0\\
b_{A \parallel} &= (0,0,-\frac{1}{2},\frac{2}{3}) \,,
\end{align}
which lead to the same GS term as deduced above.

The field theoretic interpretation along the lines of our general discussion in the previous section is sightly more tricky due to the gauging of part of the flavour group and exemplifies the remark in footnote \ref{footnote-gauging}:
The maximal non-abelian flavour group for 4 hypermultiplets in a ${\bf 2}$ of $SU(2)$ is $SO(8)$, viewing them as $8$ half-hypers in the ${\bf 8}_v$.
An $SU(3)$ subgroup of this $SO(8)$ is gauged in the present model, and the relevant branching is
\bea
SO(8) \rightarrow &U(4)_a  &\rightarrow  [SU(3) \times U(1)_{c_1}] \times U(1)_a \\
{\bf 8}_v \rightarrow    &{\bf 4}_1 + c.c. &\rightarrow {\bf 3}_{(1_{c_1}, 1_a)} + {\bf 1}_{(-3_{c_1},1_a)}  + c.c. \,.
\eea
We can ignore the complex conjugate as separate states by treating the fields again as full hypermultiplets, rather than half-hypers. In this interpretation, $U(1)_a$ appears as a `diagonal' $U(1)$.
The $SU(3)$ factor is identified as part of the gauge group.

On the other hand, the maximal flavour group acting on $6$ hypermultiplets in the ${\bf 3}$ of the gauge group $SU(3)$ is $U(6)_b$. A subgroup of $SU(2)$ is gauged, and in the present model
the commutant is further broken to an $SU(3)_F$ subgroup at the non-abelian level. The  branching rule for the first step is
\bea
U(6)_b &\rightarrow& [SU(2) \times SU(4)_F \times U(1)_{c_2}] \times U(1)_b \\
{\bf 6}_{1_b} &\rightarrow& ({\bf 2}, {\bf 1})_{(2_{c_2}, 1_b)} + ({\bf 1}, {\bf 4})_{(-1_{c_2}, 1_b)} \,,
\eea
followed by
\bea
SU(4)_F &\rightarrow& SU(3)_F \times U(1)_{c_3} \\
{\bf 4} &\rightarrow&{\bf 3}_{1_{c_3}} + {\bf 1}_{-3 c_3} \,.
\eea
While the Cartan $U(1)_{c_i}$ are manifestly free of mixed cubic $SU(3)$ ABJ anomalies (recall that $SU(2)$ is always free of cubic anomalies), the anomaly free linear combination of the diagonal $U(1)_a $ and $U(1)_b$ is generated by
\bea
T_m = 2 T_a - T_b \,.
\eea
In terms of these, the $U(1)$ flavour charges of the model, (\ref{mp1})--(\ref{mp4}), suggest writing the flavour generator as 
\bea
T_F = - \frac{1}{2} T_m + \left(  -\frac{1}{2} T_{c_1} + \frac{5}{12} T_{c_2} + \frac{1}{4} T_{c_3} \right) \,.
\eea
The maximal global symmetry of the model is
\bea
G_F = SU(4)_F \times U(1)_m \,,
\eea
even though in the present geometric realisation only an $SU(3)_F \times U(1)_F$ subgroup is manifest.

\subsection{Local model: $U(1)_F$ charged conformal matter  }\label{sec_localmodel}

In this section, we study constraints on the possible completions of a local model with a globally defined $U(1)$ gauge symmetry.
Rather than constructing a compact base $B_2$ which contains a particular 7-brane curve configuration, as we did in the previous two examples, we will use the anomaly cancellation conditions to restrict the possible abelian charges which the fields could carry in a global completion.
 We will combine these anomaly constraints with input from the structure of a putative fibration with an extra section, even without constructing a globally consistent model over a compact base.

As an example consider a local base $B_2$ which contains a configuration of shrinkable curves $C_1$, $C_2$, $C_3$ with intersection matrix
\bea \label{Cijconfiguation}
C_i \cdot C_j = \left( \begin{matrix}  -1 & 1 & 0 \\ 1 & -3 & 1 \\ 0 & 1 & -1  \end{matrix} \right) \,.
\eea
This configuration represents a so-called  conformal matter theory \cite{DelZotto:2014hpa}. For instance, it arises by blowing up the intersection point of two 7-branes with gauge algebra $\mathfrak{e}_6$. In this case, the chain of curves $C_1 - C_2  - C_3$ is sandwiched between two such curves with a corresponding $\mathfrak{e}_6$ enhancement in the Weierstrass model over $B_2$ \cite{Heckman:2013pva}.
According to the general results of \cite{Morrison:2012np}, the minimal, non-Higgsable non-abelian gauge algebra along the curve $C_2$ of self-intersection number $-3$  is $\mathfrak{su}(3)$ along $C_2$, while the (-1)-curves $C_1$ and $C_3$ carry trivial gauge algebra. In such a  non-Higgsable configuration there is no localised charged matter at the intersection of the curves.

The possible non-abelian gauge enhancements  of the above curve configuration beyond the non-Higgsable one have been classified in \cite{Heckman:2015bfa}. 
As a simple example, consider the configuration with gauge algebra
\bea \label{so8model1}
C_1: \emptyset, \qquad C_2: \mathfrak{so}(8), \qquad C_3: \emptyset  \,.
\eea
The cancellation of the pure $\mathfrak{so}(8)$ and mixed $\mathfrak{so}(8)$-gravitational anomalies uniquely determines the 
charged spectrum of this model to be given by $N_{\bf R}$ hypermultiplets in representation ${\bf R}$ with
\bea
N_{ {\bf 8}_{\rm vect}} = 1, \qquad N_{ {\bf 8}_{\rm s}} = 1, \qquad N_{ {\bf 8}_{\rm c}} = 1 \,.
\eea
These are localised at the intersection of $C_2$ with the residual component of the discriminant of the Weierstrass model over $B_2$ and away from the intersection points of $C_2$ with $C_1$ and $C_3$. 
The appearance of an equal number of hypermultiplets in the representation ${\bf 8}_{\rm vect}$, ${\bf 8}_{\rm s}$ and ${\bf 8}_{\rm c}$ is in fact a general feature of any model with  $\mathfrak{so}(8)$ algebra and follows from the subtle factorization properties of the underlying Weierstrass or Tate model \cite{Grassi:2011hq,Esole:2017qeh}.

In field theory, the maximal global symmetry acting on a single hypermultiplet in a real representation  ${\bf 8}$ of $\mathfrak{so}(8)$ is $Sp(1)$ (see e.g. \cite{Bertolini:2015bwa}). 
When we realize this model in an F-theory construction, we therefore expect to obtain at best abelian flavour symmetries $U(1)_A$ which can be interpreted as (linear combinations of) the Cartan $U(1)$s of the maximal 
$Sp(1) \times Sp(1) \times Sp(1)$ flavour group. We would like to study which constraints we can place on possible global completions of the model in which such a $U(1)$ is realized as a globally defined gauge symmetry.

From the perspective of the elliptic fibration, constructing such an abelian gauge symmetry corresponds to tuning the Weierstrass model such that it acquires an extra rational section $S_A$. In a global setup this engineers a $U(1)_A$ gauge symmetry with height-pairing $b_A$, which becomes a global flavour  symmetry upon decoupling gravity. In fact, from the general discussion at the end of \autoref{subsec_fieldexp} we know that the $U(1)$ flavour symmetry is contained as a Cartan in the non-abelian flavour symmetry  $Sp(1) \times Sp(1) \times Sp(1)$ because the model contains no hypermultiplets in a complex representation of the gauge group.
Suppose the  hypermultiplets in the various representations ${\bf 8}$ of $\mathfrak{so}(8)$ acquire $U(1)_A$ charges
$q_{ {\bf 8}_{\rm vect}}$,  $q_{ {\bf 8}_{\rm s}}$, $q_{ {\bf 8}_{\rm c}} $.
As we have shown, it is guaranteed that even in the decoupling limit the mixed $\mathfrak{so}(8) - U(1)_A$ anomaly is consistently cancelled.
With the help of
\bea
{\rm tr}_{ {\bf 8}_{\rm vect}} F^2 = {\rm tr}_{ {\bf 8}_{\rm s}} F^2 = {\rm tr}_{ {\bf 8}_{\rm c}} F^2 \equiv {\rm tr} F^2
\eea
this translates into the constraint 
\bea \label{so8U1mixed}
C_2 \cdot b_A=  2 \,( q^2_{ {\bf 8}_{\rm vect}} + q^2_{ {\bf 8}_{\rm s}} + q^2_{ {\bf 8}_{\rm c}}  )   \,,
\eea
where we have used $b_\kappa = C_2$, $\lambda_\kappa = 2$ and $A_{\rm vect} =1$.
We recall that the Shioda map takes the general form (\ref{shioda}), which we can analyse even without specifying a full global completion of the  model. 
Let us label the rational curves in the fiber over $C_2$ corresponding to the nodes  in the affine $\mathfrak{so}(8)$ Dynkin diagram as $\mathbb P^1_i$, $i= 0, 1, 2, 3, 4$. Here $\mathbb P^1_0$ refers to the affine node, $\mathbb P^1_3$ is the central node with multiplicity $2$ and $\mathbb P^1_1$, $\mathbb P^1_2$, $\mathbb P^1_4$ represent the remaining nodes with multiplicity 1. The $\mathfrak{so}(8)$ Cartan matrix $C_{ij}$, $i=1, \ldots, 4$, and its inverse take the form
\bea
 C_{ij} = \left( \begin{matrix}  2 & 0 & -1 & 0 \\
                                 0 & 2 & -1 & 0                       \\   
                                  -1 & -1 & 2 & -1                       \\    
                                 0 & 0 & -1 & 2             \end{matrix}  \right), \qquad 
 (C^{-1})^{ij} = \left( \begin{matrix}  1 & \frac{1}{2} & 1 & \frac{1}{2}  \\
                                   \frac{1}{2} & 1 & 1 & \frac{1}{2}                        \\   
                                  1 & 1 & 2 & 1                       \\    
                                   \frac{1}{2}  & \frac{1}{2} & 1 & 1             \end{matrix}  \right)
                                  \,.
                                 \eea
The extra section $S_A$ must intersect the full fiber over $C_2$ precisely once. There are only two qualitatively different patterns of  intersection numbers $\pi_{i A} = S_A \cdot \mathbb P^1_i$, $i= 1, \ldots, 4$ compatible with this requirement. If $S_A$ intersects the affine node $\mathbb P^1_0$, the intersection numbers  $\pi_{i A} = 0$ for $i = 1, \ldots, 4$ and hence there arise no  $\mathfrak{so}(8)$ correction terms in the Shioda map $\sigma_A$, (\ref{shioda}), and the height pairing $b_A$,
\bea
{\rm Model \, \,   I}: && \sigma_A = S_A-Z-\pi^{-1} (\pi_\ast((S_A-Z)\cdot Z)), \\
&&  b_A = 2 \bar K + 2 \pi_\ast(S_A \cdot  Z) \,.
\eea
Alternatively, $S_A$ can intersect any of the curves $\mathbb P^1_i$, $i=1,2,4$, but not $\mathbb P^1_3$, which has multiplicity $2$. Without loss of generality we can take the intersected node to be $\mathbb P^1_1$
and hence find for
\bea
{\rm Model \, \, II}: && \sigma_A = S_A-Z-\pi^{-1} (\pi_\ast((S_A-Z)\cdot Z)) + (E_1 + \frac{1}{2} E_2 + E_3 + \frac{1}{2} E_4),  \label{ModelIIsigmaA}\\
                             &&  b_A = 2 \bar K + 2 \pi_\ast(S_A \cdot  Z) - C_2 \,,
\eea
where $E_i$ are the $\mathfrak{so}(8)$ resolution divisors fibered over $C_2$.
Applying Riemann-Roch, (\ref{rr}), to the $(-3)$-curve $C_2$ yields 
$\bar K \cdot C_2 = - 1$, and the anomaly condition (\ref{so8U1mixed}) translates into the constraint
\[
2 \,( q^2_{ {\bf 8}_{\rm vect}} + q^2_{ {\bf 8}_{\rm s}} + q^2_{ {\bf 8}_{\rm c}}  )  = 2 \,  \pi_\ast (S_A \cdot Z) \cdot C_2   +   \begin{dcases*}
-2 &  Model I \\
+1 &  Model II
\end{dcases*}
\]

Up to this point we have not made any assumption about the explicit realisation of the Weierstrass model with an extra rational section $S_A$.
According to \cite{Morrison:2012ei},  
a large class of elliptic fibrations with one extra rational section can be expressed as a hypersurface in a ${\rm Bl}_1 \mathbb P_{112}[4]$ fibration over the base $B_2$.\footnote{This is known not to be the most general conceivable elliptic fibration with an extra section \cite{Klevers:2014bqa,Cvetic:2015ioa,Klevers:2016jsz,Raghuram:2017qut}, and it would be interesting to determine if more exotic possibilities lead to different results in the present context.}
The model is determined as the hypersurface
\bea
 \mathfrak{c}_0 w^4 s^3 +  \mathfrak{c}_1 w^3 s^2 x + \mathfrak{c}_2 w^2 s x^2 + \mathfrak{c}_3 w x^3 = y^2 s + \mathfrak{b}_0 x^2 y +  \mathfrak{b}_1  y w s x +  \mathfrak{b}_2 w^2 s^2 y  \,,
 \eea
 where $[ w : x : y]$ are homogeneous coordinates of  $\mathbb P_{112}[4]$, blown up to ${\rm Bl}_1 \mathbb P_{112}[4]$ with blow-up divisor $s$. The locus divisor 
 \bea
 S_A: s=0
 \eea
is an extra rational section.
The base polynomials $\mathfrak{b}_i$ and $\mathfrak{c}_i$ are of degree
\bea \label{degreesMP}
&& [\mathfrak{b}_0] =  2 \bar K - \beta, \quad [\mathfrak{b}_1] =  \bar K, \quad [\mathfrak{b}_2] =  \beta, \\
&&  [\mathfrak{c}_0] =  2 \beta,  \quad [\mathfrak{c}_1] =  \bar K + \beta, \quad [\mathfrak{c}_2] = 2  \bar K, \quad [\mathfrak{c}_3] =  3 \bar K - \beta \,,
\eea
where $\beta$ is an effective base divisor class 
\bea
 \beta \leq  2 \bar K \,.
\eea
Over a maximally generic base $\beta$ can be chosen such that the degree of all (holomorphic) polynomials $\mathfrak{b}_i$ and $\mathfrak{c}_i$ is non-negative. Crucially for us, the intersection of the extra section $S_A$ with the zero-section depends on $\beta$,
\bea
 \pi_\ast (S_A \cdot Z) =   {\mathfrak b}_0, \qquad [{\mathfrak b}_0] = 2 \bar K - \beta  \,.
\eea
Note that the extremal choice $\beta = 2 \bar K$ reduces the model to the $U(1)$ restricted Tate model \cite{Grimm:2010ez}, for which  $\pi_\ast (S_A \cdot Z)  = 0$.

Non-abelian gauge enhancements over certain loci on $B_2$ are engineered by specifying the explicit form of the polynomials $\mathfrak{b}_i$ and $\mathfrak{c}_i$. For the ${\rm Bl}_1 \mathbb P_{112}[4]$ fibration at hand, this has first been exemplified in \cite{Mayrhofer:2012zy, Borchmann:2013jwa,Borchmann:2013hta} and studied systematically in \cite{Kuntzler:2014ila}.
Depending on the non-abelian gauge algebra on a curve in class $[W]$, $\beta$ will be subject to a bound 
\bea
2 \bar K \geq \beta  \geq k [W]  \,, \qquad k \geq 0 \,
\eea
to ensure holomorphicity of all base polynomials. Over a specific base such as the one containing a curve configuration (\ref{Cijconfiguation}) further constraints on $\beta$ may arise in order for the fibration to exist.

A realisation of Model II along $C_2$ with locus coordinate $C_2: \gamma = 0$ is given by specifying the vanishing orders  \cite{Kuntzler:2014ila}
\bea
{\mathfrak c}_0 = {\mathfrak c}_{0,3}  \gamma^3,   \qquad 
{\mathfrak c}_1 = {\mathfrak c}_{1,2}  \gamma^2,   \qquad   
{\mathfrak c}_2 = {\mathfrak c}_{2,1}  \gamma,      \qquad 
{\mathfrak c}_3 = {\mathfrak c}_{3,1}  \gamma,   \qquad
{\mathfrak b}_1 = {\mathfrak b}_{1,1}  \gamma,    \qquad 
{\mathfrak b}_2 = {\mathfrak b}_{2,1}  \gamma \,.
\eea
More precisely, this corresponds to a so-called canonical model, in which there are no further non-trivial relations between the $ {\mathfrak c}_{i,j}$ and $ {\mathfrak b}_{k,l}$. 
Over a generic base, the discriminant of the fibration then takes the form
\bea
\Delta = \gamma^6 \left(  {\mathfrak b}_0^2  {\mathfrak b}_{2,1}^2  {\mathfrak c}_{2,1}^2   ({\mathfrak b}_0{\mathfrak b}_{21,}+ {\mathfrak c}_{2,1})^2 + {\cal O}(\gamma) \right) \,.
\eea
This indicates four enhancement loci at the intersection of $\gamma =0$ with any of the four factors of  ${\mathfrak b}_0^2  {\mathfrak b}_{2,1}^2  {\mathfrak c}_{2,1}^2 ({\mathfrak b}_0{\mathfrak b}_{21,}+ {\mathfrak c}_{2,1})^2$.
Even without constructing  a concrete base $B_2$ containing the configuration (\ref{Cijconfiguation}), anomaly cancellation allows us to find important necessary conditions which such a global model has to comply with.
First, non-abelian anomaly cancellation shows that the intersection of $C_2$ with one of the above four polynomials defining the matter loci must be trivial in order to arrive at the required number of precisely three hypermultiplets in the  ${\bf 8}$ representations.
In view of the degrees of the polynomials (\ref{degreesMP}) this is only possible if one of the following two possibilities occurs,
\bea
&& {\rm Case \,\,  A)} \quad  \beta \cdot C_2 = -2     \qquad \Longrightarrow   \qquad \pi_\ast (S_A \cdot Z) \cdot C_2 = 0   \\
&& {\rm Case \,\,   B)}  \quad  \beta \cdot C_2 = -3    \qquad   \Longrightarrow  \qquad  \pi_\ast (S_A \cdot Z) \cdot C_2 = 1 \,.
\eea
From the form (\ref{ModelIIsigmaA}) of the correction terms in $\sigma_A$
 one concludes 
the charges $q_i$ of the localised hypermultiplets can at best be half-integer. This is because the charges are computed by the intersection numbers of $\sigma_A$ with localised fibral curves in codimension-two, and the only source of fractions for these charges are the half-integer coefficients of the resolution divisors $E_i$ in  $\sigma_A$.
It is then easy to see that the only possible configurations of charges for Model II compatible with mixed $\mathfrak{so}(8)-U(1)_A$ anomaly cancellation are 
\bea \label{possibilities}
&& {\rm Model \, II, \, \, Case \,  A)}   \qquad (q_{ {\bf 8}_{\rm vect}},q_{ {\bf 8}_{\rm s}}, q_{ {\bf 8}_{\rm c}}) \in \{(\frac12,\frac12,0),  (\frac12,0,\frac12),  (0,\frac12,\frac12) \}   \\
&& {\rm Model \, II, \, \,  Case \,  B)}   \qquad (q_{ {\bf 8}_{\rm vect}},q_{ {\bf 8}_{\rm s}}, q_{ {\bf 8}_{\rm c}}) \in \{(\frac12,\frac12,1),  (\frac12,1,\frac12),  (1,\frac12,\frac12) \}  \,.  
\eea
This is the normalization where extra $U(1)$ charged hypermultiplets transforming as $\mathfrak{so}(8)$ singlets, which necessarily occur away from the configuration of shrinkable curves, have charge $1$ and $2$ \cite{Morrison:2012ei}. 
The inclusion of a $U(1)_A$ flavour symmetry hence necessarily breaks the triality between the three hypermultiplets in representations ${\bf 8}_{\rm vect}$, ${\bf 8}_{\rm s}$, and ${\bf 8}_{\rm c}$, but this by itself is not an inconsistency.
Whether a global extension of this local configuration on a base $B_2$ together with a rational section really exists is a different question and requires explicit construction. 

Similarly, Model I can be  constructed in canonical form by imposing the vanishing orders \cite{Kuntzler:2014ila}
\bea
{\mathfrak c}_0 = {\mathfrak c}_{0,4}  \gamma^4,   \qquad 
{\mathfrak c}_1 = {\mathfrak c}_{1,2}  \gamma^2,   \qquad   
{\mathfrak c}_2 = {\mathfrak c}_{2,1}  \gamma,      \qquad 
{\mathfrak b}_1 = {\mathfrak b}_{1,1}  \gamma,    \qquad 
{\mathfrak b}_2 = {\mathfrak b}_{2,2}  \gamma^2 
\eea
together with the extra factorization condition
\bea
4 {\mathfrak c}_{1,2} {\mathfrak c}_{3} - {\mathfrak c}_{2,1}^2 = \tau^2 \,.
\eea
The discriminant can be computed as 
\bea
\Delta  =  \gamma^6 \left( {\mathfrak c}_{1,2}^2 \,  {\mathfrak c}_{3}^2 \,  \tau^2    + \cal{O}(\gamma) \right) \,.
\eea
Non-abelian anomaly cancellation gives the following constraints over a base with configuration (\ref{Cijconfiguation}) and $\mathfrak{so}(8)$ enhancement over the (-3) curve $C_2$:
The hypermultiplets sit at the intersection points of $C_2$ with  $\tau$, ${\mathfrak c}_{3}$ and ${\mathfrak c}_{1,2}$. Precisely 3 copies of hypermultiplets are found if either 
$\beta \cdot C_2 = -4$ or $\beta \cdot C_2 = -3$ or $\beta \cdot C_2 = -5$. In the first case, all three types of different loci carry one hypermultiplet, whereas in the second case the intersection of $C_2$ with either ${\mathfrak c}_{3}$ or ${\mathfrak c}_{1,2}$ is trivial, while the respective other intersection locus carries $2$ hypermultiplets.

The possible charge assignments consistent with the cancellation of mixed anomalies are
\bea
{\rm Model \, I:}   && \beta \cdot C_2 = - 5: \qquad \qquad (q_{ {\bf 8}_{\rm vect}},q_{ {\bf 8}_{\rm s}}, q_{ {\bf 8}_{\rm c}}) \in \{(1,1,0),  (0,1,1),  (1,0,1) \} \\
       && \beta \cdot C_2 = - 4: \qquad \qquad (q_{ {\bf 8}_{\rm vect}},q_{ {\bf 8}_{\rm s}}, q_{ {\bf 8}_{\rm c}}) \in \{(1,0,0),  (0,1,0),  (0,0,1) \} \\
      &&    \beta \cdot C_2 = - 3: \qquad \qquad (q_{ {\bf 8}_{\rm vect}},q_{ {\bf 8}_{\rm s}}, q_{ {\bf 8}_{\rm c}}) \in \{(0,0,0) \} \,.
\eea
In the latter case, the $U(1)_A$ gauge symmetry acts trivially on the $\mathfrak{so}(8)$ states.
As pointed already, if we take a decoupling limit, the $U(1)_A$ with the charges determined above becomes a global symmetry which in fact is a linear combination of the Cartan $U(1)$s of the maximal field theoretic global symmetry $Sp(1) \times Sp(1) \times Sp(1)$.

\section{Conclusions and Prospects} \label{sec_concl}

In this work we have explored $U(1)$ flavour symmetries in 6d $N=(1,0)$ SCFTs. Such theories can be obtained starting from $N=(1,0)$ supergravity theories and taking the limit of decoupling gravity. 

In the first part  of this paper we have shown that the structure of 6d gauge anomaly cancellation implies that the $U(1)$ gauge symmetries of a supergravity theory inevitably turn into a flavour symmetry once gravity is decoupled. A clear geometric interpretation has then been given to supergravity theories with a string/F-theory embedding specified by a {\it compact} elliptic Calabi-Yau three-fold: The strength of the $U(1)$ interaction is known to be inversely proportional to the volume of a complex curve in the compact base $B_2$, given by the so-called height pairing of a rational section.
Our supergravity analysis hence implies that the height pairing cannot be contractible as otherwise the $U(1)$ would remain as a gauge theory after decoupling gravity. 
Motivated by this prediction from supergravity we have studied the contractibility properties of a general height pairing purely from the perspective of geometry. We have proven its non-contractibility without relying on the physics of anomalies. The proof uses that the anti-canonical divisor $\bar K$ of any F-theory base $B_2$ is an effective piece of the height pairing together with the fact that $\bar K$ is non-contractible. 

Having established the fate of $U(1)$ symmetries in the decoupling limit from a geometric and a supergravity perspective, we have proceeded to analyse abelian flavour symmetries of 6d $N=(1,0)$ SCFTs. More precisely, we have focused on the flavour group as far as its action on the 6d hypermultiplet matter is concerned. Up to a few subtleties \cite{Ohmori:2015pia}, the SCFT flavour symmetries can be explored by investigating the flavour symmetries of the tensor branch theories. The latter enjoy a geometric description in terms of the decoupling limit of F-theory on an elliptic Calabi-Yau three-fold. 
In view of the richness of possible constructions of abelian gauge theories in {\it compact} F-theory, one might at first sight worry that the possibility of having abelian flavour symmetries leads to a proliferation of possibilities in the list of SCFTs \cite{Heckman:2013pva,DelZotto:2014hpa,Heckman:2015bfa,Heckman:2015ola,Bhardwaj:2015xxa}. However, we have argued that the abelian flavour symmetries can be determined already locally from the perspective of the maximal global flavour group: If we assign to $N$ hypermultiplets in a complex representation of the gauge group a maximal flavour group $U(N)$, rather than $SU(N)$, the flavour symmetries can be understood as anomaly free combinations of the diagonal $U(1)$ factors. This is in fact in perfect agreement with previous results on the origin also of abelian gauge groups in F-theory versus Type IIB theory \cite{Krause:2012yh,MayorgaPena:2017eda}. In a given F-theory realisation, these may obtain admixtures from the Cartan subgroup of the maximal non-abelian part of the flavor symmetry.
We have tested this proposal in concrete examples in which the local flavour symmetry has a global completion to a gauge symmetry.

An important physics aspect that has only briefly been mentioned in this work is that of discrete symmetries. 
Abelian discrete symmetries  are encoded  in the F-theory geometry by torsion homology \cite{Camara:2011jg,Mayrhofer:2014laa} or, less directly, by  a non-trivial Tate-Shafarevich group of a genus-one fibered Calabi-Yau manifolds  \cite{Braun:2014oya,Morrison:2014era, Anderson:2014yva,Klevers:2014bqa,Garcia-Etxebarria:2014qua,Mayrhofer:2014haa,Mayrhofer:2014laa,Cvetic:2015moa}. 
At the level of supergravity theories, discrete gauge symmetries may arise from Higgsing global $U(1)$s.
In addition, the so-called massive $U(1)$s, whose massless linear combinations we have identified in \autoref{subsec_fieldexp} as the field theoretic origin of the (non-Cartan) flavour $U(1)$s, may survive as discrete global symmetries.
Recently, \cite{Anderson:2018heq} has presented SCFTs with discrete flavour charges.
It would be interesting to generalize our analysis of the continuous $U(1)$ flavour symmetries to a discrete version in this context, to which we wish to come back in the near future. 

Furthermore, while the focus of this work has been placed on 6d theories, it would be worth analyzing their lower-dimensional analogues, e.g., 4d and 2d theories. To begin with, it would be interesting to carefully study the fate of a $U(1)$ gauge symmetry upon decoupling gravity from such a lower-dimensional supergravity theory, be it embedded in a string theoretic  framework or not. The physical constraints on the $U(1)$s are particularly easy to deal with in 6d, where the purely abelian one-loop anomalies are quartic and thereby lead to a definite sign for the GS contribution. Similarly, the geometric study for the contractibility property of height pairing is much simpler for a two-fold base $B_2$ than for their higher-dimensional cousins. 
We leave a further investigation of $U(1)$s for the lower-dimensional theories to future work.

\section*{Acknowledgements}

We thank A. Grassi, A. Hanany, C. Lawrie, W. Lerche, L. Lin, C. Mayrhofer, P. Oehlmann, C. Reichelt, S. Sch\"afer-Nameki, M. Walters and F. Xu for helpful discussions.
This work was partly funded by TR 33 `The Dark Universe'.


\begin{appendix}

\section{The Mordell-Weil group and abelian gauge symmetries in F-theory}  \label{reviewMW}

In this appendix we review the origin of abelian gauge symmetries in the Mordell-Weil group of rational sections of an F-theory elliptic fibration (\ref{ellfibration1}).

\subsection{Rational sections and $U(1)_A$ symmetries}

A rational section $s$ is a rational (i.e. meromorphic) map from $B_2$ to $\hat Y_{3}$, which assigns to each generic point $b$ on the base a unique point $s(b)$ in the fiber over $b$.
 Such maps form a group called Mordell-Weil group 
 \bea
 MW(\pi) \simeq \IZ^{\oplus r} \oplus MW(\pi)_{\rm tor} \,,
 \eea
 which is a finitely generated abelian group with respect to the natural arithmetic of the elliptic fibration $\pi$. 
Since each section intersects a generic elliptic fiber in precisely one point, the group law on the space of sections can be defined by addition of the image points on the generic elliptic fiber. 
The zero element of this abelian group is represented by the zero-section $s_0$, whose existence distinguishes an elliptic fibration from a genus-one fibration without section and which maps each point on the base to the zero-point on the generic elliptic fiber. The number of independent non-torsional sections $r$ is called the rank of the Mordell-Weil group.

The relation between the abelian gauge group factors and $ MW(\pi)$ originates in the observation that each independent non-torsional section $s_A$ defines an independent divisor class 
 \bea
 S_A := {\rm div}(s_A) \in H_4(\hat Y_3)
 \eea
on $\hat Y_{3}$. This is because $s_A(B_2)$ defines an embedding of the base into the full fibration, which is a 4-cycle on $\hat Y_3$ whose associated divisor class we denote by $S_A$. 
When we compactify M-theory on $\hat Y_{3}$, expanding the 3-form gauge potential $C_3$ in terms of the Poincar\'e dual 2-form gives rise to a vector field in the M-theory effective action in $\mathbb R^{1,4}$, which is related to the $U(1)_A$ gauge potential in the dual F-theory modulo a few subtleties which we now recall:

Suppose a divisor class $\mathfrak{D}$ on $\hat Y_{3}$ gives rise to a 1-form potential in the M-theory effective action in $\mathbb R^{1,4}$ by expanding
\bea
C_3 = A_{\mathfrak{D}} \wedge [\mathfrak{D}] + \ldots \,,  
\eea 
where $[\mathfrak{D}] \in H^2(\hat Y_3)$ is the Poincar\'e dual 2-form associated to $\mathfrak{D}$. 
In order for the 1-form potential $A_{\mathfrak{D}}$ to lift to a 7-brane $U(1)$ gauge potential in the dual F-theory, the divisor class $\mathfrak{D}$ must satisfy 
two constraints known as the transversality conditions\footnote{The symbol $\cdot$ denotes the intersection product in homology. To avoid confusion, we will sometimes specify on which space it acts by a subscript e.g. of the form $\cdot_{\hat Y_3}$ for the intersection product in $H_\ast(\hat Y_3)$.}
\bea \label{transverse1}
\mathfrak{D} \cdot Z \cdot D_{\alpha_i} = 0, \qquad \mathfrak{D} \cdot  D_{\alpha} \cdot  D_{\beta} = 0 \,
\eea
in terms of the zero-section divisor $ Z= {\rm div}(s_0)$
and for any divisor pulled back from the base,
 \bea \label{Dalpha}
 D_{\alpha} = \pi^{-1} (D^{\rm b}_{\alpha}), \qquad D^{\rm b}_{\alpha} \in H_{2}(B_2) \,.
 \eea
 In the presence of non-abelian gauge group factors, an extra condition must be imposed, as reviewed at the beginning of  \autoref{6g_general}. Up until Appendix \ref{Correctionsnonab} we assume that no such non-abelian gauge groups are present.

The two conditions (\ref{transverse1}) require a modification of the divisor $S_A$ in order for it to give rise to a $U(1)_A$ gauge group factor in F-theory. The correct linear combination of divisors is in fact given by \cite{Grimm:2010ez,Park:2011ji}
\beq\label{shioda1}
\sigma(s_A) = S_A-Z-\pi^{-1} \left(\pi_\ast((S_A-Z)\cdot Z)\right)  \in H_4(\hat Y_4) \,.
\eeq
The expression makes use of the pushforward map $\pi_\ast$ induced by the projection $\pi$ of the elliptic fibration,
\bea \label{pushforwardmap}
 \pi_\ast:  H_{k}(\hat Y_{3}) \rightarrow H_{k}(B_2) \,. 
\eea
According to the projection formula for all $\omega \in H_k(B_2)$, $\gamma \in H_{4-k}(\hat Y_3)$,
\bea
\pi^{-1}(\omega) \cdot_{\hat Y_3} \gamma = \omega \cdot_{B_2} \pi_\ast(\gamma) \,.
\eea
With the help of this formula, it is clear that $\sigma(S_A)$ satisfies (\ref{transverse1}).
In arithmetic geometry, the map (\ref{shioda1}) is known as the Shioda homomorphism~\cite{shioda, wazir}, 
\beq  \label{Shioda-homgen}
\sigma: MW(\pi) \to H_{4}(\hat Y_{3}) \ , 
\eeq
which is a homomorphism from the Mordell-Weil group $MW(\pi)$ to the homology group $H_{4}(\hat Y_{3})$.

In the dual M-theory the expansion of the 3-form gauge potential
\bea
C_3 = \sum_{A=1}^r A^A \wedge [\sigma(s_A)] + \ldots
\eea
gives rise to abelian gauge potentials $A^A$ which lift to the gauge potentials of the gauge group factor $U(1)_A$ in F-theory.

\subsection{Gauge couplings and the height pairing in absence of non-abelian gauge groups} \label{app_Gaugecoupl}

The abelian gauge kinetic term in M-theory follows by dimensional reduction of the kinetic term of $C_3$ as
\bea \label{Skin1}
S_{\rm kin} = - \frac{2\pi}{2} \int_{{\cal M}^{1,10}}  dC_3 \wedge \ast d C_3 = - \frac{2\pi}{2}  f_{A B}  \int_{\mathbb R^{1,4}} dA^A \wedge \ast d A^B 
\eea 
with
\bea
 f_{A B} = \int_{\hat Y_{3}}   [\sigma(s_A)] \wedge \ast  [\sigma(s_B)] \,. 
\eea
On the Calabi-Yau 3-fold $\hat Y_3$ with K\"ahler form $J_{\hat Y_3}$ this expression can be further expressed as \cite{Candelas:1990pi},
\bea \label{fABM}
f_{AB} =  -  \int_{\hat Y_3}  J_{\hat Y_3} \wedge [\sigma(s_A)] \wedge [\sigma(s_B)]  +      \left( \int_{\hat Y_3} J_{\hat Y_3}^2 \wedge [\sigma(s_A)] \right)   \left( \int_{\hat Y_3} J_{\hat Y_3}^2 \wedge [\sigma(s_B)]   \right)    \left(  \int_{\hat Y_3}   J_{\hat Y_3}^3  \right)^{-1} \,.
\eea
While this determines the couplings of the gauge potentials in the M-theory effective action, the gauge kinetic terms in the dual F-theory are considerably simpler.
This is because in the F-theory limit the volume of the fiber is taken to zero while at the same the volume of the base is  rescaled.
More precisely, if we expand $J_{\hat Y_3}$ in a basis of $H^2(\hat Y_3)$ as
\bea
J_{\hat Y_3} = J  +  t^0  [Z] + t^A \pi^{-1} [S_A] \,, \qquad  J = t^\alpha \,  \pi^{\ast} (\omega_\alpha) \,,
\eea 
with $\omega_\alpha$ a basis of $H^2(B_2)$, then the 6d F-theory limit corresponds to the scaling \cite{Bonetti:2011mw}
\bea
t^\alpha  \rightarrow \epsilon^{-1/2} \,  t^\alpha, \qquad t^0 \rightarrow  \epsilon \,  t^0, \qquad  t^A \rightarrow  \epsilon \,  t^A
\eea
and taking $\epsilon \rightarrow 0$. 
The only surviving terms in this limit involve $J$, the part of the K\"ahler form 
pulled back from the base $B_2$.
Since $\sigma(s_A)$ and $\sigma(s_B)$ satisfy (\ref{transverse1}), the second term in (\ref{fABM}) vanishes in the F-theory limit and we are left with
\bea
f_{AB} \rightarrow \hat f_{AB}    =  - \int_{\hat Y_3} \pi^* J \wedge [\sigma(s_A)] \wedge   [\sigma(s_B)] =  \int_{B_2} J \wedge [ - \pi_\ast (\sigma(s_A) \cdot   \sigma(s_B)) ] \,.
\eea
This is the kinetic matrix governing the kinetic terms of the abelian gauge factors in F-theory,
\bea
S_{\rm kin} |_{\rm F-theory}  =     - \frac{2\pi}{2}  \hat f_{A B}  \int_{\mathbb R^{1,5}} dA^A \wedge \ast d A^B  \,.
\eea

Note that the intersection product $\sigma(s_A) \cdot  \sigma(s_B)$ defines an element in $H_{2}(\hat Y_{3})$ and pushing this onto the base gives
a divisor class
\bea\label{heightp}
\Ub_{AB}:=  - \pi_\ast (\sigma(s_A) \cdot  \sigma(s_B)) \in H_{2}(B_2).
\eea
 
The object $\Ub_{AB}$ is known in arithmetic geometry as the \emph{height pairing} of the sections $s_A$ and $s_B$ and will play a central role in our analysis.

The evaluation of the height pairing is well-known in the mathematics (and in the F-theory \cite{Braun:2011zm,Krause:2011xj,Grimm:2011fx,Morrison:2012ei}) literature and proceeds with the help of the intersection numbers of the elliptic fibration $\hat Y_3$ as follows. 
Let us abbreviate the pullback divisor in the Shioda homomorphism (\ref{shioda1a}) as  
\bea
D_A = \pi^{-1} \left(\pi_\ast((S_A-Z)\cdot Z) \right) \in H_4(\hat Y_3)
\eea
such that
\begin{equation}
\begin{split}
\Ub_A&=-\pi_\ast \left((S_A-Z-D_A)\cdot(S_A-Z-D_A) \right) \cr \label{U}
&= -\pi_\ast (S_A \cdot S_A)-\pi_\ast (Z \cdot Z) + 2 \pi_\ast (S_A\cdot Z) - \pi_\ast({D_A}\cdot D_A) + 2 \pi_\ast((S_A-Z)\cdot D_A) \,.
\end{split}
\end{equation}
In order to further simplify Eq.~\eqref{U}, we first note that $S_A$, being a section to the fibration, obeys the intersection relations
\bea
S_A \cdot S_A \cdot  D_{\alpha} &=& - \pi^{-1}(\bar K) \cdot S_A \cdot  D_{\alpha} \ ,\label{brS} \\
S_A\cdot D_\alpha \cdot D_\beta &=&   D^{\rm b}_{\alpha} \cdot_{B_2}  D^{\rm b}_{\beta}  \label{oguiso}
\eea
for any pullback divisor of the form 
\bea \label{Dalpha1}
 D_{\alpha} = \pi^{-1} (D^{\rm b}_{\alpha}), \qquad D^{\rm b}_{\alpha} \in H_{2}(B_2) \,.
 \eea

The same relations   hold for the zero section $Z$.\footnote{The reader is kindly referred to Ref.~\cite{Anderson:2016ler} for more details, as well as for applications to finding sections to an elliptic fibration in the context of F-theory.} 
Here $\bar K \equiv - K$ is the anti-canonical divisor of the base $B_2$.
On the other hand, a triple product involving only pullback divisors vanishes,
\bea
D_\alpha \cdot D_\beta \cdot D_\gamma = 0 \ .
\eea
It is then straightforward to see that the last two terms in the heigh pairing~\eqref{U} vanish and the expression simplifies as
\beq \label{hp}
\Ub_A=2\bar K + 2\pi_\ast(S_A \cdot Z) \,.
\eeq

\subsection{Corrections from non-abelian gauge group factors} \label{Correctionsnonab}

 Let us finally discuss the modifications of the Shioda homomorphism in the presence of non-abelian gauge groups.
 As discussed at the beginning of  \autoref{6g_general}, in this case the singular elliptic fibration is resolved by the inclusion of the resolution divisors (\ref{Eik}), whose fibers are the curves  $\mathbb P^1_{i_\ik}$.
 In the dual M-theory compactification on $\hat Y_3$, M2-branes wrapping the fibral curves $\mathbb P^1_{i_\ik}$ give rise to gauge bosons whose Cartan charges are given by the (negative of the) positive simple roots of the Lie algebra $\mathfrak{g}_\ik$. The resolution divisors $E_{i_\ik}$ give rise to the Cartan $U(1)$s via expansion of the M-theory 3-form $C_3$.
 In view of the physical origin of the non-abelian gauge bosons as wrapped M2-branes along the curves $\mathbb P^1_{i_\ik}$, their Cartan charges are given by (minus one times) the intersection numbers with $E_{i_\ik}$, i.e. 
  \bea \label{Cartan-def}
E_{i_\kappa} \cdot \mathbb P^1_{j_\lambda} = - \delta_{\kappa \lambda}  \, C_{i_\kappa j_\kappa} 
\eea
 with $C_{i_\ik j_\ik}$   the Cartan matrix  of the non-abelian gauge group $G_\ik$.
 
 In the presence of non-abelian gauge group factors $G_\ik$, it is desirable to normalize the non-Cartan factors $U(1)_A$ associated with the Mordell-Weil group such that the non-abelian gauge bosons are uncharged under it. The constraint we need to impose is hence that   
\bea \label{DP1=0}
\sigma(s_A) \cdot   \mathbb P^1_{i_\ik} = 0 \,.
\eea
In order to satisfy this constraint in addition  to (\ref{transverse1}), 
 the Shioda homomorphism~\eqref{Shioda-homgen} necessarily acquires additional contributions from the resolution divisors and is given in total by
\beq\label{shioda}
\sigma(s_A) = S_A-Z-\pi^{-1} (\pi_\ast((S_A-Z)\cdot Z))+ \sum_\ik \sum_{i_\ik}  \ell_{A}^{i_\ik}   E_{i_\ik} \ . 
\eeq 
The correction terms involving the resolution divisors $E_{i_\ik}$ are required only to implement (\ref{DP1=0}) and involve the coefficients 
$\ell_{A}^{i_\ik} \in \mathbb Q$. 
These are easily computed as
\bea \label{lAjik}
\ell_{A}^{i_\ik}  = \pi_{A j_\ik} (C^{-1})^{j_\ik i_\ik}
\eea
in terms of the intersection numbers
\bea
\pi_{A i_\ik} = S_A \cdot \mathbb P^1_{i_\ik}
\eea
and the Cartan matrix $C_{i_\ik j_\ik}$ of the non-abelian gauge group $G_\ik$, which arises because of (\ref{Cartan-def}).

\section{6d $N=(1,0)$ Anomalies} \label{sec_anomalies}

The low-spin massless representations of six-dimensional (1,0) supersymmetry are labeled by their $SU(2)\times SU(2)$ representations. Namely,
\begin{align}\nonumber
\text{Gravity multiplet}&: \quad(1,1)+2(\frac{1}{2},1)+(0,1)\\\nonumber
\text{Tensor multiplet}&: \quad(1,0)+2(\frac{1}{2},0)+(0,0)\\\nonumber
\text{Vector multiplet}&: \quad(\frac{1}{2},\frac{1}{2})+2(0,\frac{1}{2})\\\nonumber
\text{Hypermultiplet}&: \quad2(\frac{1}{2},0)+4(0,0)
\end{align}
All of these multiplets contain chiral fields which means that they contribute to the one-loop anomaly. In particular, the contribution of each multiplet to the anomaly polynomial is
\bea
I^{\rm gravity} &=& -\frac{273}{5760} ({\rm tr} R^4 + \frac54({\rm tr}R^2)^2)  + \frac{9}{128}({\rm tr}R^2)^2 \\
I^{\rm tensor} &=& \frac{29}{5760} ({\rm tr} R^4 + \frac54({\rm tr}R^2)^2)  -\frac{1}{128}({\rm tr}R^2)^2 \\
I^{\rm vector} &=&- d_G \frac{1}{5760} ({\rm tr} R^4 + \frac54({\rm tr}R^2)^2) -\frac{1}{24} {\rm Tr} F^4 + \frac{1}{96} {\rm Tr} F^2 {\rm tr}R^2  \\
I^{\rm hyper} &=&d_\rho \frac{1}{5760} ({\rm tr} R^4 + \frac54({\rm tr}R^2)^2)  + \frac{1}{24}{\rm tr}_{\rho} F^4  - \frac{1}{96} {\rm tr}_\rho F^2 {\rm tr}R^2 
\eea
where $G$ is the gauge group, including Abelian factors.
In the above expressions, $\rm Tr$ is the trace in the adjoint representation and ${\rm tr}_\rho$ corresponds to the trace in the representation $\rho$ of the gauge group $G$. $d_G$ and $d_\rho$ denote the dimensions of the gauge group and of the representation $\rho$. It is useful to recall the following group-theoretic factors: 
\bea
{\rm tr}_R F^2 &=& A_R {\rm tr} F^2 \\
{\rm tr}_R F^3 &=& E_R {\rm tr} F^3 \\ 
{\rm tr}_R F^4 &=& B_R {\rm tr} F^4 + C_R ({\rm tr} F^2)^2 \ ,
\eea
where ${\rm tr}$ denotes the trace in the fundamental representation of $G$.

When the total one-loop anomaly polynomial factorizes, it can be cancelled by the addition of a Green-Schwarz-Sagnotti-West term,
\begin{align}
S^{GS}=-\frac{1}{2}\int \Omega_{\alpha\beta}B^\alpha\wedge X_4^\beta\,.
\end{align}
Indeed, such a term is not gauge-invariant and gives the following contribution to the anomaly polynomial
\begin{align}
I_8^{GS}=-\frac{1}{32}\Omega_{\alpha\beta}X_4^\alpha \wedge X_4^\beta\,,
\end{align}
with $X_4^\alpha$ is given in \eqref{X4alpha}.
The total anomaly polynomial is then
\begin{align}
I_8=I_8^{\rm 1-loop}+I_8^{GS}
\end{align}
where $I_8^{\rm 1-loop}$ contain the contribution of every massless fundamental field in the tensor branch.

We should mention that when the gauge group contains a $U(1)$ factor with field strength $F=dA$, there is an additional GSSW term that can be added to the action,
\begin{align}
\widetilde S^{GS}=\int \phi X_6\,,
\end{align}
where $\phi$ is a scalar and $X_6$ is a six-form. This term may cancel a contribution to the anomaly polynomial of the form $F\wedge X_6$ as long as the kinetic form for the scalar is schematically $(d\phi+A)^2$. Since this mechanisms makes the $U(1)$ massive we will not discuss it further.

Cancellation of anomalies for a gauge group $G=\prod_{A=1}^r U(1)_A\times \prod_{\kappa}G_\kappa$, i.e.~$I_8=0$, imposes that
\begin{align}
273 &= H - V + 29T\\
a\cdot a &= 9 - T\\
a\cdot b_\kappa &=\frac{1}{6}\lambda_\kappa(A_{\rm Adj_\kappa}-\sum_I\mathcal M_I^\kappa A^I_\kappa)\\
0 &= B_{\rm Adj_\kappa}-\sum_I \mathcal M_I^\kappa B^I_\kappa\\
b_\kappa\cdot b_\kappa &= \frac{1}{3}\lambda_\kappa^2(\sum_I \mathcal M_I^\kappa C_\kappa^I-C_{\rm Adj_\kappa})\\
b_\kappa\cdot b_\mu &= \lambda_\kappa\lambda_\mu\sum_I\mathcal M_I^{\kappa \mu} A_\kappa^I A_\nu^I\\
a\cdot b_{AB} &= -\frac{1}{6}\sum_I\mathcal M_I q_{IA}q_{IB}\\
0 &= \sum_I\mathcal M_I^\kappa E^I_\kappa q_{IA} \label{cubicmixed}\\ 
\frac{b_\kappa}{\lambda_\kappa}\cdot b_{AB} &= \sum_I \mathcal M_I^\kappa A_\kappa^I q_{IA}q_{IB}\\ 
b_{AB}\cdot b_{CD}+b_{AC}\cdot b_{BD}+b_{AD}\cdot b_{BC}&= \sum_I\mathcal M_I q_{IA} q_{IB} q_{IC}q_{ID}  \ . 
\end{align}
Here the index $I$ runs over the irreducible representations of the non-abelian gauge group that appear in the spectrum. $\mathcal M_I^{\kappa}$ and $\mathcal M_I^{\kappa\mu}$ denote the number of $G_\kappa$ and $G_{\kappa}\times G_{\mu}$ representations in $I$. Finally, $H$, $V$ and $T$ denote the total number of hyper, vector and tensor multiplets.

\end{appendix}

\newpage
\bibliography{papers}
\bibliographystyle{custom1}

\end{document}